\documentclass[printer]{aa}
\usepackage[utf8]{inputenc}
\usepackage[varg]{txfonts}
\usepackage{color}
\usepackage{amsmath}
\usepackage{graphicx}
\usepackage[english]{babel}
\usepackage{siunitx}
\usepackage{float}
\usepackage{url}
\usepackage{longtable}
\usepackage{fancyhdr}
\usepackage{natbib}
\usepackage[normalem]{ulem}

\title{The CARMENES search for exoplanets around M dwarfs}
\subtitle{The warm super-Earths in twin orbits around the mid-type M dwarfs \\ Ross~1020 (GJ~3779) and LP~819-052 (GJ~1265)}
\titlerunning{Two twin super-Earths orbiting the M dwarfs GJ~3779 and GJ~1265}
\author{
        R.~Luque\inst{1,2}
        \and
        G.~Nowak\inst{1,2}
        \and
        E.~Pall{\'e}\inst{1,2}
        \and
        D.~Kossakowski\inst{3}
        \and
        T.~Trifonov\inst{3}
        \and
        M.~Zechmeister\inst{4}
        \and
        V.~J.~S.~Béjar\inst{1,2}
        \and
        C.~Cardona~Guillén\inst{1,2}
        \and
        L.~Tal-Or\inst{4,14}
        \and
        D.~Hidalgo\inst{1,2}
        \and
        I.~Ribas\inst{5,6}
        \and
        A.~Reiners\inst{4}
        \and
        J.~A.~Caballero\inst{7}
        \and
        P.~J.~Amado\inst{8}
        \and
        A.~Quirrenbach\inst{9}
        \and
        J.~Aceituno\inst{10}
        \and
        M.~Cortés-Contreras\inst{7}
        \and
        E.~Díez-Alonso\inst{11}
        \and
        S.~Dreizler\inst{4}
        \and
        E.~W.~Guenther\inst{12}
        \and
        T.~Henning\inst{3}
        \and
        S.~V.~Jeffers\inst{4}
        \and
        A.~Kaminski\inst{9}
        \and
        M.~Kürster\inst{3}
        \and
        M.~Lafarga\inst{5,6}
        \and
        D.~Montes\inst{7}
        \and
        J.~C.~Morales\inst{5,6}
        \and
        V.~M.~Passegger\inst{13}
        \and
        J.~H.~M.~M.~Schmitt\inst{13}
        \and
        A.~Schweitzer\inst{13}
        }
        
\institute{
        Instituto de Astrof{\'i}sica de Canarias, 
        E-38205 La Laguna, Tenerife, Spain\\ \email{rluque@iac.es}
        \and
        Dept.~Astrof{\'i}sica, Universidad de La Laguna, 
        E-38206 La Laguna, Tenerife, Spain
        \and
        Max-Planck-Institut für Astronomie, Königstuhl 17, D-69117 Heidelberg, Germany
        \and
        Institut für Astrophysik, Georg-August-Universität, Friedrich-Hund-Platz 1, D-37077 Göttingen, Germany
        \and
        Institut de Ciències de l'Espai (ICE,CSIC), Campus UAB, c/ de Can Magrans s/n, E-08193 Bellaterra, Barcelona, Spain
        \and
        Institut d'Estudis Espacials de Catalunya (IEEC), E-08034 Barcelona, Spain
        \and 
        Centro de Astrobiología (CSIC-INTA), ESAC campus, Camino Bajo del Castillo s/n, E-28692 Villanueva de la Cañada, Madrid, Spain
        \and
        Instituto de Astrofísica de Andalucía (IAA-CSIC), Glorieta de la Astronomía s/n, E-18008 Granada, Spain
        \and
        Landessternwarte, Zentrum für Astronomie der Universtät Heidelberg, Königstuhl 12, D-69117 Heidelberg, Germany
        \and 
        Centro Astronómico Hispano-Alemán (CSIC-MPG), Observatorio Astronómico de Calar Alto, Sierra de los Filabres, E-04550 Gérgal, Almería, Spain
        \and 
        Departamento de Astrofísica y Ciencias de la Atmósfera, Facultad de Ciencias Físicas, Universidad Complutense de Madrid, E-28040 Madrid, Spain
        \and 
        Thüringer Landessternwarte Tautenburg, Sternwarte 5, D-07778 Tautenburg, Germany
        \and 
        Hamburger Sternwarte, Gojenbergsweg 112, D-21029 Hamburg, Germany
        \and 
        School of Geosciences, Raymond and Beverly Sackler Faculty of Exact Sciences, Tel Aviv University, Tel Aviv 6997801, Israel
           }
\date{Received 14 May 2018 / Accepted 16 Oct 2018}

\abstract{
We announce the discovery of two planetary companions orbiting around the low mass stars Ross~1020 (GJ~3779, M4.0V) and LP~819-052 (GJ~1265, M4.5V). The discovery is based on the analysis of CARMENES radial velocity observations in the visual channel as part of its survey for exoplanets around M dwarfs. In the case of GJ~1265, CARMENES observations were complemented with publicly available Doppler measurements from HARPS. The datasets reveal one planetary companion for each star that share very similar properties: minimum masses of $8.0\pm0.5$\,M$_{\oplus}$ and $7.4\pm0.5$\,M$_{\oplus}$ in low-eccentricity orbits with periods of $3.023\pm0.001$\,d and $3.651\pm0.001$\,d for GJ~3779~b and GJ~1265~b, respectively. The periodic signals around three days found in the radial velocity data have no counterpart in any spectral activity indicator. Besides, we collected available photometric data for the two host stars, which confirm that the additional Doppler variations found at periods around 95\,d can be attributed to the rotation of the stars. The addition of these planets in a mass-period diagram of known planets around M dwarfs suggests a bimodal distribution with a lack of short-period low-mass planets in the range of 2--5\,M$_{\oplus}$. It also indicates that super-Earths (> 5 M$_{\oplus}$) currently detected by radial velocity and transit techniques around M stars are usually found in systems dominated by a single planet.
}


\begin{document}

\maketitle

\section{Introduction}

The search for exoplanets has become a prominent research field in the past twenty years, particularly the detection of rocky planets in the habitable zones of their parent stars. The radial velocity (RV) technique has been successfully applied to detect such companions around M dwarfs given the relatively low mass of these stars \citep{Marcy98,Rivera05,Udry07,Bonfils13,ProxCen}. M stars account for three quarters of all stars known within 10\,pc of our Solar System \citep{2016AAS...22714201H} and show an occurrence rate in average of more than two planets per host star \citep{Dress15,2016MNRAS.457.2877G}. 

Despite their high potential for finding rocky planets, M dwarfs pose various observational difficulties. On the one hand, low-mass stars emit more flux in the near-infrared than in the optical. On the other hand, chromospheric variability and activity cycles may produce changes in the spectral line profiles, which mimic a Doppler shift. These periodic variations induce signals in the RV data that could be wrongly interpreted as of planetary origin \citep[e.g.][]{2001A&A...379..279Q,2014Sci...345..440R,Sarkis18}. Since stellar jitter can reach amplitudes of a few $\mathrm{m\,s^{-1}}$, detecting low-signal companions generally requires a considerable number of observations \citep{2011MNRAS.412.1599B}. However, since the amplitude of an activity-related signal is expected to be wavelength-dependent, a successful program searching for exoplanets around M dwarfs has to tackle these difficulties by observing simultaneously in the widest possible wavelength range, especially covering the reddest optical wavelengths, and designing an observing strategy that ensures numerous and steady observations covering a long timespan.

The CARMENES search for exoplanets around M dwarfs accomplishes these requirements in its Guaranteed Time Observation (GTO) M dwarf survey, which began in January 2016 \citep{Reiners17}. Ross~1020 (GJ~3779) and LP~819-052 (GJ~1265) are two mid-type M dwarfs monitored as part of this project. Their RVs indicate the presence of two super-Earths in short-period orbits of the order of 3\,d. Section~\ref{sec:stars} summarizes the basic information of the host stars. In Section~\ref{sec:data}, the radial velocity observations for each star are presented.  The analysis of the RV data is explained in Section~\ref{sec:fit}, while in Section~\ref{sec:discussion} we discuss the results from the Keplerian fit and the location of the two planets in a mass-period diagram.

\section{Stellar parameters} \label{sec:stars}

\begin{table}
\centering
{\renewcommand{\arraystretch}{1.1}
 \renewcommand{\tabcolsep}{0.005\textwidth}
 \small 
\caption{Stellar parameters for GJ~3779 and GJ~1265}. \label{tab:stars}
\begin{tabular}{lccr}
\hline\hline
\noalign{\smallskip}
Parameter 				& GJ~3779 & GJ~1265 		& Ref.\tablefootmark{a} \\ 
\noalign{\smallskip}
\hline
\noalign{\smallskip}
\multicolumn{4}{c}{Main identifiers and coordinates}\\
\noalign{\smallskip}
Name                    & Ross~1020     & LP~819-052        &       \\
Karmn                    & J13229+244     & J22137-176        &       \\
$\alpha$				& 13:22:56.74		& 22:13:42.78	   & 2MASS 	\\
$\delta$				& +24:28:03.4		& -17:41:08.2	   & 2MASS 	\\
\noalign{\smallskip}
\multicolumn{4}{c}{Spectral type and magnitudes}\\
\noalign{\smallskip}
SpT				& M4.0~V		& M4.5~V	   & PMSU 	\\ 
$G$ [mag]					& $11.6266 \pm 0.0008$	& $12.0807 \pm 0.0006$	   & {\it Gaia DR2} 	\\
$J$ [mag]					& $8.728 \pm 0.02$	& $8.955 \pm 0.03$	   & 2MASS 	\\
\noalign{\smallskip}
\multicolumn{4}{c}{Kinematics}\\
\noalign{\smallskip}
$d$ [pc]					& $13.748 \pm 0.011$	& $10.255 \pm 0.007$	   & {\it Gaia DR2} 	\\
$\mu_{\alpha}\cos\delta$ [$\mathrm{mas\,yr^{-1}}$]      & $-615.95 \pm 0.13$	& $856.89 \pm 0.12$	   & {\it Gaia DR2} 	\\
$\mu_{\delta}$ [$\mathrm{mas\,yr^{-1}}$]     & $-865.13 \pm 0.10$	& $-306.30 \pm  0.11$	   & {\it Gaia DR2} 	\\
$V_r$ [$\mathrm{km\,s^{-1}}]$        & $-19.361$	& $-24.297$	   & Rei18 	\\
\noalign{\smallskip}
\multicolumn{4}{c}{Photospheric parameters}\\
\noalign{\smallskip}
$T_{\mathrm{eff}}$ [K]			& $3324 \pm 51$	& $3236 \pm 51$	   & This work\tablefootmark{b} 	\\
$\log g$				& $5.05 \pm 0.07$	& $5.09 \pm 0.07$	   & This work\tablefootmark{b} 	\\
{[Fe/H]}				& $0.00 \pm 0.16$	& $-0.04 \pm 0.16$   & This work\tablefootmark{b} 	\\
\noalign{\smallskip}
\multicolumn{4}{c}{Derived physical parameters}\\
\noalign{\smallskip}
$L$ [10$^{-5}$\,L$_{\odot}$]		& $867 \pm 11$	& $364 \pm 5$	   &  This work	\\
$R$ [R$_{\odot}$]			& $0.281 \pm 0.010$		& $0.192 \pm 0.007$	   & This work 	\\
$M$ [M$_{\odot}$]			& $0.27 \pm 0.02$	& $0.178 \pm 0.018$   & This work 	\\
\noalign{\smallskip}
\multicolumn{4}{c}{Stellar rotation}\\
\noalign{\smallskip}
$v \sin i$ [$\mathrm{km\,s^{-1}}$]			& $< 2.0$		& $< 2.0$	   & Rei18      \\
$P_{\mathrm{rot}}$ [d]			& $95 \pm 5$		& $> 70$	   & This work\\
\noalign{\smallskip}
\hline
\end{tabular}}
\tablefoot{
\tablefoottext{a}{{\it References}. 2MASS: \citet{2MASS}; PMSU: \citet{PMSU}; {\it Gaia DR2}: \citet{GaiaDR2}; Rei18: \citet{Reiners17}.}
\tablefoottext{b}{Estimated as in \citet{Passegger18}.}
}
\end{table}

The basic information of the host stars GJ~3779 and GJ~1265 is presented in Table~\ref{tab:stars}. Both targets exhibit similar properties and their values are consistent with the literature. They are mid-type M dwarfs that kinematically belong to the Galactic thin disc \citep{CC16}, with GJ~1265 being part of the young population. Their photospheric metallicities are compatible with solar values \citep{Neves14,Newton14,Passegger18}. They are thought to be inactive, with no emission in H$\alpha$ \citep{Jeffers18} and only very faint X-ray emission for the case of GJ~1265 \citep{3XMM}. The photospheric parameters of the stars were determined as in \citet{Passegger18} using the latest grid of PHOENIX-ACES model spectra \citep{PHOENIX-ACES} and the method described in \citet{Passegger16}. We determined the radii $R$ using Stefan-Boltzmann's law, the spectroscopic $T_{\mathrm{eff}}$ as in \citet{Passegger18}, and the luminosities $L$ by integrating the photometric stellar energy distribution collected for the CARMENES targets (Caballero et al. 2016) with the Virtual Observatory Spectral energy distribution Analyser \citep{Bayo08}. The stellar masses were derived from the linear $M$--$R$ relation presented by \citet{Schweitzer18}. These values are the same using instead empirical mass--luminosity relations such the ones presented by \citet{Delfosse00} and \citet{Benedict16}. The details on the calculation of the physical parameters were described in \citet{Reiners18} and will be reported in further detail by \citet{Schweitzer18}.

The star GJ~3779 (Ross~1020, J13229+244) is a high proper motion star classified as M4.0~V by \citet{PMSU}. It resides in the Coma Berenices constellation, located at a distance of $d = 13.748\pm0.011\,\mathrm{pc}$ \citep{GaiaDR2} with an apparent magnitude in the $J$ band of 8.728\,mag \citep{2MASS}. Using CARMENES data, \citet{Reiners17} determined its absolute radial velocity to be $V_r = -19.361\,\mathrm{km\,s^{-1}}$ and a Doppler broadening upper-limit of $v \sin i < 2\,\mathrm{km\,s^{-1}}$.

GJ~1265 (LP~819-052, J22137-176) is also a high proper motion star at a distance of $d = 10.255\pm0.007\,\mathrm{pc}$ \citep{GaiaDR2} in the Aquarius constellation. Its apparent magnitude in the $J$ band is 8.955\,mag \citep{2MASS}, and it is approaching the Solar System with an absolute radial velocity of $V_r = -24.297\,\mathrm{km\,s^{-1}}$ \citep{Reiners17}. The star exhibits a luminosity in X-rays of $\log L_\mathrm{X} = 26.1\pm0.2\,\mathrm{erg\,s^{-1}}$, measured with the \textit{XMM-Newton} observatory \citep{3XMM}. Therefore, we can estimate the rotational period of the star to be of the order of $100\,\mathrm{d}$ following the $L_\mathrm{X}$--$P_{\mathrm{rot}}$ relation proposed by \citet{Reiners14}.

\section{Observations} \label{sec:data}

The CARMENES instrument, installed at the 3.5\,m Calar Alto telescope at the Calar Alto Observatory in Spain, consists of a simultaneous dual-channel fiber-fed high-resolution spectrograph covering a wide wavelength range from $\SI{0.52}{\micro\metre}$ to $\SI{0.96}{\micro\metre}$ in the visual (VIS) and from $\SI{0.96}{\micro\metre}$ to $\SI{1.71}{\micro\metre}$ in the near-infrared (NIR), with spectral resolutions of $R = 94\,600$ and $R = 80\,400$, respectively \citep{CARMENES}. The performance of the instrument and its ability to discover exoplanets around M dwarfs using the RV technique has been already demonstrated by \citet{Trifonov18}, \citet{Reiners18} and \citet{Kaminski18}.

The survey contains 104 observations for GJ~3779 in the VIS channel covering a timespan of 820\,d, with typical exposure times of 1800\,s and median internal RV precision of $\sigma_{\mathrm{VIS}} = 1.7\,\mathrm{m\,s^{-1}}$. In the case of GJ~1265, 87 observations were acquired in the visible, covering a timespan of 782\,d, typical exposure times of 1800\,s and median internal RV precision of $\sigma_{\mathrm{VIS}} = 1.8\,\mathrm{m\,s^{-1}}$. The NIR median internal RV precisions are considerably higher, $\sigma_{\mathrm{NIR}} = 10.4\,\mathrm{m\,s^{-1}}$ and $\sigma_{\mathrm{NIR}} = 6.7\,\mathrm{m\,s^{-1}}$ for GJ~3779 and GJ~1265, respectively. As a result, we only use the RVs from the VIS channel in our analysis.

Doppler measurements, together with several spectral indicators useful to discuss the planetary hypothesis, were obtained with the SERVAL program \citep{SERVAL} using high signal-to-noise templates created by co-adding all available spectra of each star. They are corrected for barycentric motion \citep{WE14}, as well as for secular acceleration, which is important for stars with high proper motions \citep{Zechmeister09}. In addition, the RVs were corrected by means of an instrumental nightly zero-point (NZP), which was calculated from a subset of RV-quiet stars observed each night whose RV standard deviation is less than $10\,\mathrm{m\,s^{-1}}$. This correction is described in \citet{Trifonov18}. For each spectrum, we also computed the cross-correlation function (CCF) using a weighted mask by co-adding the stellar observations themselves. Fitting a Gaussian function to the final CCF, we determined the radial velocity, FWHM, contrast, and bisector velocity span. The methodology regarding how the CCF is computed for our CARMENES spectra is described in \citet{Reiners18}. 

Moreover, for our southern target GJ~1265, we further included 11 publicly available observations from HARPS \citep{HARPS} taken between July 2003 and July 2014. We used SERVAL to derive their radial velocities and obtained a median internal uncertainty of $\sigma_{\mathrm{HARPS}} = 5.6\,\mathrm{m\,s^{-1}}$. The individual radial velocities for both stars are listed in Tables~\ref{tab:gj3779_RVs} and \ref{tab:gj1265_RVs}.

\begin{figure}
\centering
\includegraphics[width=\hsize]{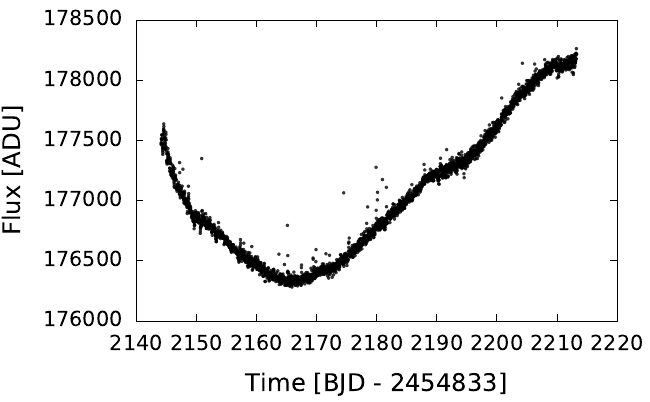}
\caption{{\it K2} light curve of GJ~1265. The photometric variability of the star is longer than the 69\,d baseline of the observations.} \label{fig:gj1265_lc}
\end{figure}

In addition to the Doppler measurements, we also collected available photometry for both targets with the goal of determining the rotational period. We analyzed photometric time series for GJ~3779 from The MEarth Project \citep{MEarth,MEarth12} from January 2009 to July 2010 in the RG715 filter. For GJ~1265 (EPIC~205913009) we employed photometric time series from the \textit{K2} space mission during its Campaign 3 \citep{K2} from November 2014 to February 2015. Fig.~\ref{fig:gj1265_lc} shows the photometric variability of GJ~1265. Assuming that the signal is close to sinusoidal we derive a rotational period of $\sim 95$\,d, which agrees with the $\sim 100$\,d rotational period estimated from the X-ray luminosity presented in Sect.~\ref{sec:stars}. However, due to the mix of periodic and stochastic behaviour of stellar variability and the fact that the {\it K2} timespan does not cover a full phase of such period, we can only conclude that GJ~1265 is a slowly rotating star with a period longer than 70\,d.

\section{Data analysis} \label{sec:fit}

\subsection{GJ~3779 b} \label{subsec:gj3779b}

\begin{figure}
\centering
\includegraphics[width=\hsize]{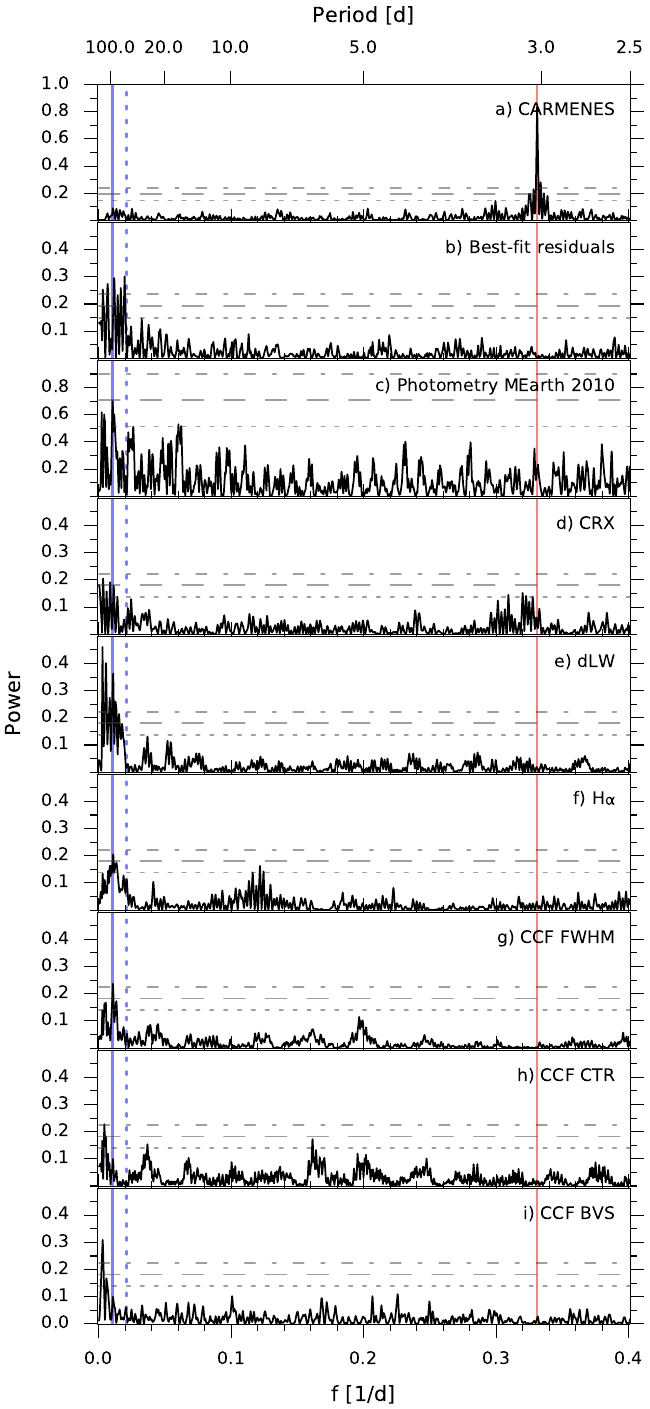}
\caption{GLS periodograms for GJ~3779 CARMENES RVs (a) and its residuals (b) after removing the prominent peak at $f = 0.3307\,\mathrm{d^{-1}}$ ($P = 3.023\,\mathrm{d}$), marked in red. Panel c shows the periodogram of a photometric campaign of MEarth data from 2010. The highest peak at $f_\mathrm{rot} = 0.0105\,\mathrm{d^{-1}}$ ($P = 95.2\,\mathrm{d}$) associated to the star's rotation is marked with a solid blue line and its first harmonic ($2f_\mathrm{rot}$) with a dashed blue line. Panels d--f show periodograms of the chromatic index, differential line width, and H$\alpha$ index, while panels g--i show periodograms for the cross-correlation function full-width half-maximum, contrast, and bisector velocity span. Horizontal lines show the theoretical FAP levels of 10\% (short-dashed line), 1\% (long-dashed line) and 0.1\% (dot-dashed line) for each panel.
} \label{fig:gj3779_gls}
\end{figure}

\begin{figure}
\centering
\includegraphics[width=\hsize]{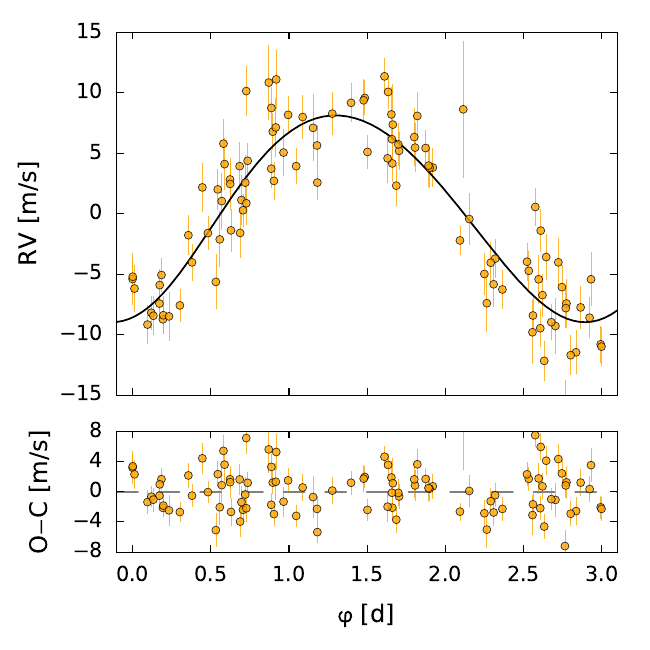}
\caption{CARMENES radial velocities and residuals phase-folded to the best Keplerian fit consistent with a $3.023\,\mathrm{d}$ period planet around GJ~3779.} \label{fig:gj3779_rv}
\end{figure}

\begin{figure}
\centering
\includegraphics[width=\hsize]{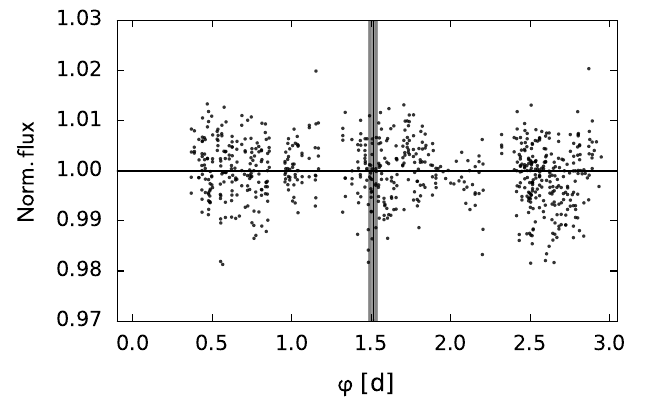}
\caption{Phase-folded light curve to the $3.023\,\mathrm{d}$ planet period of the MEarth photometric data for GJ~3779. The gray shaded area marks the expected transit duration. The central time of the expected transit is also marked with a vertical bar.} \label{fig:gj3779_transit}
\end{figure}

In Fig.~\ref{fig:gj3779_gls}, we show the generalized Lomb-Scargle periodograms \citep[GLS,][]{GLS} of different data products obtained with SERVAL and our CCF analysis. For each panel, we compute the theoretical false alarm probability (FAP) as described in \citet{GLS}, and show the 10\%, 1\%, and 0.1\% levels. In the CARMENES RVs (Fig.~\ref{fig:gj3779_gls}a) a single relevant peak with $\mathrm{FAP}=2.2\times10^{-35}$ at $f = 0.3307\,\mathrm{d^{-1}}$ ($P = 3.023\,\mathrm{d}$) is found. The next highest peak in the RVs GLS, though above $\mathrm{FAP}=10\%$, is found at $f = 0.0105\,\mathrm{d^{-1}}$ ($P = 95.24\,\mathrm{d}$), which agrees with the periodic photometric variation found in the MEarth data (Fig.~\ref{fig:gj3779_gls}c). After subtraction of the 3.023\,d periodic signal in the RV GLS, no further significant signals are found (Fig.~\ref{fig:gj3779_gls}b). 

A strong alias\footnote{For the sake of clarity we only show the full GLS periodogram up to $1\,\mathrm{d^{-1}}$ in the Appendix (Fig.~\ref{fig:gls_ext}).} of the signal discussed so far is found at a frequency of $1\,\mathrm{d^{-1}} - f = 0.6693\,\mathrm{d^{-1}}$. A periodogram of the data does not provide information about which of the two peaks is the true planetary signal. However, the peak at $1\,\mathrm{d^{-1}} - f$, i.e. $P = 1.494\,\mathrm{d}$, has less power and therefore appears less likely to be the true period. If the fitted signal is the one at $P = 1.494\,\mathrm{d}$, the $3.023\,\mathrm{d}$ peak in the RV GLS does not disappear, proving that the $P = 1.494\,\mathrm{d}$ signal is indeed an alias of the $\sim 3\,\mathrm{d}$ one. Furthermore, if we choose the true Keplerian period to be the shortest one at $\sim 1.5$\,d, the goodness of the fit gets significantly worse, changing from $\chi^2_{\nu} = 2.6$ to $\chi^2_{\nu} = 5.6$, and the derived eccentricity of the orbit becomes larger, which is unlikely compared to the period and eccentricity distributions of known exoplanets.

Taking full advantage of the spectral information provided by SERVAL and the one contained in the CCF, we further investigate possible periodic signals related with chromospheric activity of the host star that may have induced RV variations. Periodograms of the chromatic index (CRX), differential line width (dLW), and H$\alpha$ index as described in \citet{SERVAL} are shown in the panels d--f of Fig.~\ref{fig:gj3779_gls}, while full width half maximum (FWHM), contrast (CTR) and bisector velocity span (BVS) from the CCF are shown in panels g--i of Fig.~\ref{fig:gj3779_gls}. 

Only the dLW GLS shows significant peaks between 80 and 100\,d, almost coincident with those found in the photometry, the radial velocities, best-fit residuals and the CCF FWHM. We also investigated the nature of the marginal power close to $\sim 3\,\mathrm{d}$ with $\mathrm{FAP}\sim10\%$ present in Fig.~\ref{fig:gj3779_gls}d. A significant signal in the CRX at the RV period may indicate that the measured Doppler shifts are dependent on wavelength, which would refute the planetary hypothesis. However, we checked that there is no correlation in the RV--CRX scatter plot that may call into question the nature of the periodic signal at $f = 0.3307\,\mathrm{d^{-1}}$ (see left panel of Fig.~\ref{fig:rv_crx} in the Appendix). None of the other indicators show periodicity at this frequency and, hence, we conclude that the periodic signal at $3.023\,\mathrm{d}$ is due to a planetary origin and the signal at $P \sim 95\,\mathrm{d}$ is related to the rotation of the star.

To obtain the orbital parameters of the planet, we fit the RV dataset searching for the $\chi^2$ minimum using the Levenberg-Marquardt method \citep{1992nrfa.book.....P} in a Keplerian scheme. The best-fit Keplerian orbital parameters and their corresponding 1$\sigma$ uncertainties of the planet GJ~3779~b are listed in Table~\ref{tab:fit_all}. We estimated the uncertainties of the best-fit parameters using two different approaches. First, we obtained the errors given by the covariance matrix in the Levenberg-Marquardt algorithm itself. Second, using the Markov Chain Monte Carlo (MCMC) sampler \texttt{emcee} \citep{emcee} and a Keplerian model with flat priors, we can derive median and 1$\sigma$ uncertainty values from the posterior distributions. A noise term ("jitter") is modeled simultaneously during the parameter optimization following \citet{Baluev09}. The posterior distributions from the MCMC analysis are shown in Fig.~\ref{fig:gj3779_mcmc} in the Appendix.

The phased RVs and their residuals around the best fit are shown in Fig.~\ref{fig:gj3779_rv}. The abscissa values of phase-folded plot are determined by computing the difference between each Julian date and the epoch of the first RV observation, and then computing the remainder of the quotient of this difference with the orbital period. The planet candidate GJ~3779~b has a minimum mass of $8.0\pm0.5\,\mathrm{M_{\oplus}}$ orbiting at a semi-major axis of 0.026\,au, much closer to the host star than the conservative habitable-zone limits \citep[0.094--0.184\,au,][]{Kopparapu13,Kopparapu14}. The results from best-fit and MCMC analyses all agree within the uncertainties. The posterior distributions show Gaussian profiles and peak close the best-fit values from Table~\ref{tab:fit_all}.

Finally, using the photometric time series data, we looked for the possibility that GJ~3779~b transits its host star. After detrending and flattening, no significant peaks were found in the Box-fitting Least Squares (BLS) periodogram \citep{BLS}. Nevertheless, we shifted the observations to the predicted secondary transit following \citet{Kane09} and phase-folded the light curve to the 3.023\,d period of the planet. We did not find any hints of transits, as shown in Fig.~\ref{fig:gj3779_transit}. However, since the radius of the planet can be between 1--3\,$\mathrm{R_{\oplus}}$ depending on the volatile content, the depth of a hypothetical transit could be between 0.1--1.0\%. In addition, since the expected duration of the transit is approximately one hour, the uncertainties of the transit window are big and there are gaps in the phase-folded photometric data, there could still be a chance that the transit was missed or hidden in the noise of the MEarth data. Therefore, we conclude that more data would be required to rule out completely the possibility that GJ~3779~b transits its host star.

\subsection{GJ~1265 b} \label{subsec:gj1265b}

\begin{figure}
\centering
\includegraphics[width=\hsize]{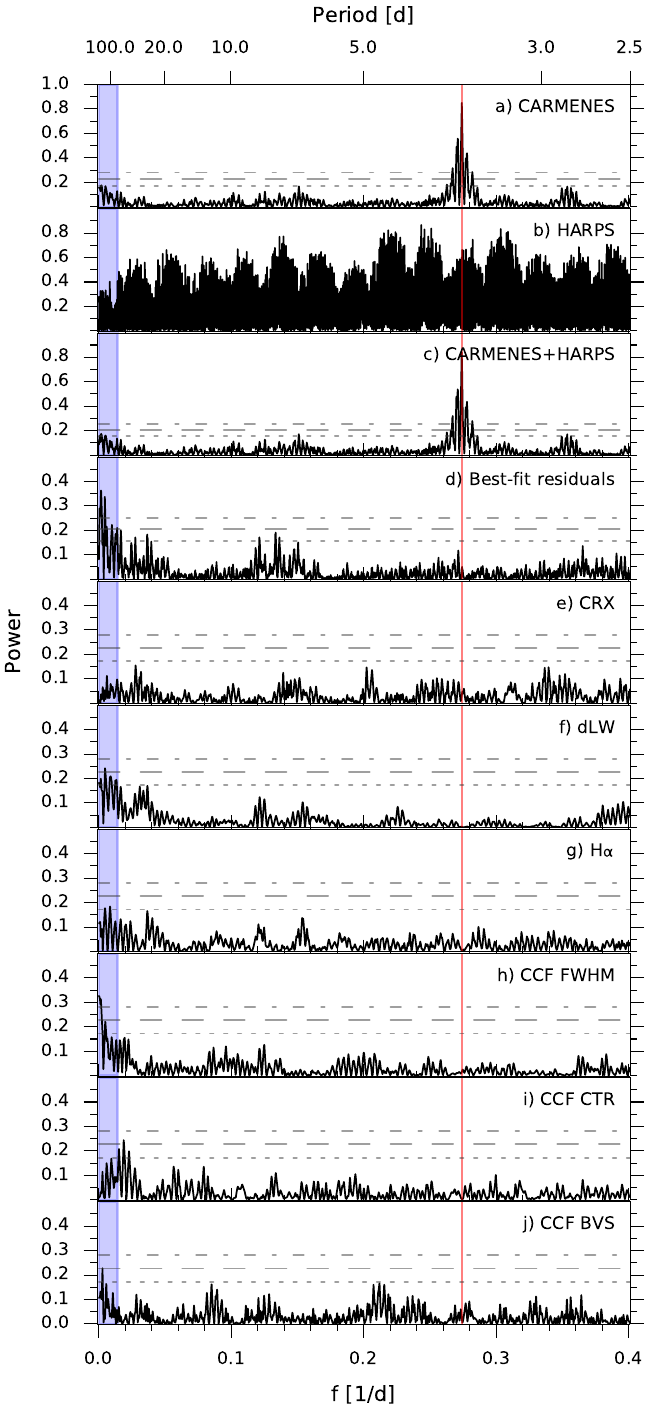}
\caption{GLS periodograms for GJ~1265 CARMENES (a), HARPS (b), and combined CARMENES+HARPS (c) data. Panel d shows the residuals after removing the prominent peak at $f = 0.2739\,\mathrm{d^{-1}}$ ($P = 3.651\,\mathrm{d}$), marked in red. The blue shaded area depicts the excluded planetary region due to stellar variability as discussed in Sect.~\ref{sec:data}. Panels e--j and horizontal lines are same as Fig.~\ref{fig:gj3779_gls}. 
 } \label{fig:gj1265_gls}
\end{figure}

\begin{figure}
\centering
\includegraphics[width=\hsize]{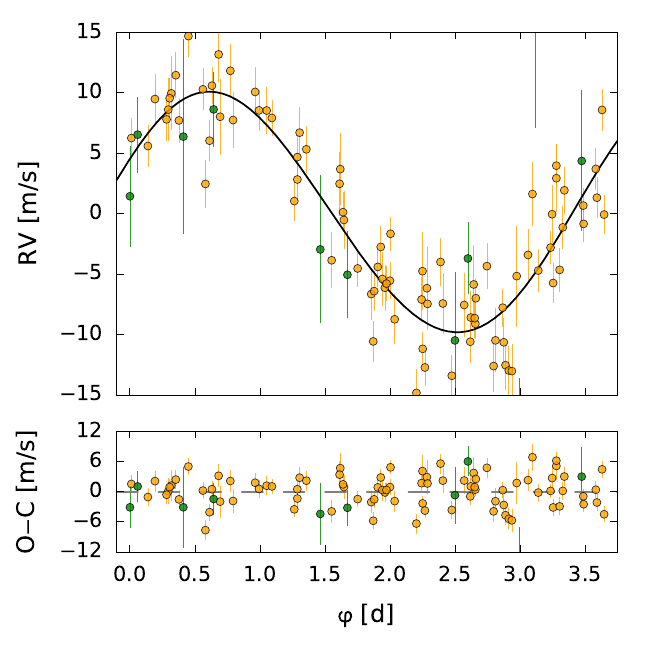}
\caption{CARMENES (orange) and HARPS (green) radial velocities and residuals phase-folded to the best Keplerian fit consistent with a $3.651\,\mathrm{d}$ period planet around GJ~1265.}  \label{fig:gj1265_rv}
\end{figure}

\begin{figure}
\centering
\includegraphics[width=\hsize]{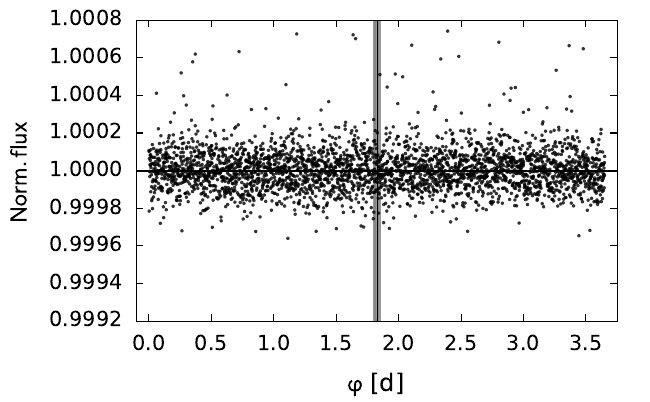}
\caption{Phase-folded light curve to the $3.651\,\mathrm{d}$ planet period of the \textit{K2} photometric data for GJ~1265. The gray shaded area marks the expected transit duration. The central time of the expected transit is also marked with a vertical bar.} \label{fig:gj1265_transit}
\end{figure}

The same approach as for the previous system has been applied for the RV analysis of GJ~1265. The first three panels of Fig.~\ref{fig:gj1265_gls} show the GLS periodograms for the CARMENES, HARPS, and combined (CARMENES+HARPS) radial velocities. A prominent single peak at $f = 0.2739\,\mathrm{d^{-1}}$ ($P = 3.651\,\mathrm{d}$) and its corresponding alias at $1\,\mathrm{d^{-1}}-f$ (Fig.~\ref{fig:gls_ext}, right panel) can be seen in the CARMENES dataset. Even though the HARPS measurements are not sufficient and too spread in time to detect the short-period signal in the GLS, when combined with the CARMENES radial velocities, the aforementioned peak increases in significance from $\mathrm{FAP}=2.1\times10^{-29}$ using only CARMENES data to $\mathrm{FAP}=1.7\times10^{-32}$ adding the HARPS RVs, also improving the frequency resolution. The nature of the peak at $1\,\mathrm{d^{-1}}-f$ has been analyzed in the same way as in Sect.~\ref{subsec:gj3779b}, concluding that it is an alias of the true Keplerian signal. Fitting a sinusoid at 3.651\,d, the GLS of the residuals (Fig.~\ref{fig:gj1265_gls}d) exhibits some power at $f < 0.012\,\mathrm{d^{-1}}$ ($P > 83.3\,\mathrm{d}$), which is compatible with the stellar variability seen in the \textit{K2} photometry (Fig.~\ref{fig:gj1265_lc}) and the predicted rotational period derived in Sect.~\ref{sec:stars}.

In the periodogram analysis of the activity indicators, no remarkable peaks have been found at the frequency of the radial velocity signal or the rotational period of the star. However, following the procedure explained in Sect.~\ref{subsec:gj3779b}, we checked the nature of the power at $\sim 5$\,d in the CRX. Like in the previous case, no significant correlation have been found between the RVs and the CRX in the CARMENES VIS data (Fig.~\ref{fig:rv_crx}, right panel), neither between HARPS RVs and the CCF parameters or H$\alpha$. Gathering all the evidence, we conclude that the periodic signal found in the CARMENES RVs can only be due to the star's reflex motion caused by a planetary companion. Therefore, we took the RVs GLS periodogram peak as initial period guess to fit a Keplerian orbit to the combined RV dataset using the Levenberg-Marquardt method. The best-fit Keplerian orbital parameters and 1$\sigma$ uncertainties of the planet candidate GJ~1265~b are listed in Table~\ref{tab:fit_all}, with their corresponding MCMC analysis as explained in Sect.~\ref{subsec:gj3779b} shown in Fig.~\ref{fig:gj1265_mcmc}.

Combined phase-folded RVs together with their best-fit are depicted in Fig.~\ref{fig:gj1265_rv}.  The rms around the best-fit is of $3.0\,\mathrm{m\,s^{-1}}$ and $8.3\,\mathrm{m\,s^{-1}}$ for the CARMENES and HARPS RVs, respectively, which corresponds to $1.5\sigma$ and $1.2\sigma$ given their internal RV uncertainties. The planet orbiting around the M dwarf star GJ~1265 shows similar properties as the one presented in Sect.~\ref{subsec:gj3779b}, with a minimum mass of $7.4\pm0.5\,\mathrm{M_{\oplus}}$ in a 3.651\,d low-eccentricity orbit at a semi-major axis of 0.026\,au. The MCMC analysis agrees with the best-fit results within the errors. 

As mentioned in Sect.~\ref{sec:data}, GJ~1265 was observed by the \textit{K2} space mission during its Campaign 3. Using this photometry information, we analyzed the light curve to look not only for periodic signals associated to the rotation of the star, but also for planetary transits. After removing the photometric variation caused by the rotation of the star, no further signals were found in the BLS periodogram. Nonetheless, we phase-folded the photometric data to the radial velocity signal at 3.651\,d to look for possible transits, but we found none at the period of the planet, as shown in Fig.~\ref{fig:gj1265_transit}.

\section{Discussion and summary} \label{sec:discussion}

\begin{table}
\centering
{\renewcommand{\arraystretch}{1.15}
\caption{Keplerian orbital parameters and $1\sigma$ uncertainties of the planet candidates.}  \label{tab:fit_all}
\sisetup{separate-uncertainty = true,table-sign-mantissa}
\begin{tabular}{lcc}
\hline\hline
\noalign{\smallskip}
\multicolumn{1}{l}{Parameter} & \multicolumn{1}{c}{Best-fit}  & \multicolumn{1}{c}{MCMC} \\
\noalign{\smallskip}
\hline
\noalign{\smallskip}
\multicolumn{3}{c}{GJ 3779 b} \\ 
\noalign{\smallskip}
\hline
\noalign{\smallskip}
$K$ [$\mathrm{m\,s^{-1}}$]  & 8.61$\,\pm\,$0.40 	    & $8.62^{+0.39}_{-0.39}$		\\ 
$P$ [d] 			        & 3.0232$\,\pm\,$0.0013   	& $3.0232^{+0.0004}_{-0.0004}$	    \\ 
$e$  				        & 0.07$\,\pm\,$0.05  	    & $0.06^{+0.05}_{-0.06}$	     \\ 
$\omega$ [deg] 		        & 225$\,\pm\,$42 	& $231^{+42}_{-54}$	    \\ 
$M$ [deg] 	                & 339$\,\pm\,$41	& $333^{+52}_{-41}$		  \\
\noalign{\smallskip}
\noalign{\smallskip} 
$\gamma_{\mathrm{VIS}}$ [$\mathrm{m\,s^{-1}}$]          & 0.85$\,\pm\,$0.52 	    	& $0.86^{+0.30}_{-0.29}$ 		  \\ 
$\sigma_{\mathrm{jitt,VIS}}$ [$\mathrm{m\,s^{-1}}$]          & 2.2\ (fixed) 	    	& $2.19^{+0.29}_{-0.29}$ 		  \\ 
\noalign{\smallskip}
\noalign{\smallskip}
$m_p\sin{i}$ [M$_{\oplus}$] & 8.0$\,\pm\,$0.5     & 		  \\ 
$a$ [au]                    & 0.026$\,\pm\,$0.001	& 		  \\ 
\noalign{\smallskip}    
\hline
\noalign{\smallskip}
\multicolumn{3}{c}{GJ 1265 b} \\ 
\noalign{\smallskip}
\hline
\noalign{\smallskip}
$K$ [$\mathrm{m\,s^{-1}}$]  & 9.89$\,\pm\,$0.47 	& $9.82^{+0.51}_{-0.52}$ 		\\ 
$P$ [d] 			        & 3.6511$\,\pm\,$0.0012   & $3.6511^{+0.0001}_{-0.0001}$	 \\ 
$e$  				        & 0.04 $\,\pm\,$0.04  	& $0.06^{+0.04}_{-0.06}$	 \\ 
$\omega$ [deg] 		        & 282$\,\pm\,$72 	    & $293^{+24}_{-25}$	\\ 
$M$ [deg] 	                & 19$\,\pm\,$74	        & $9^{+30}_{-27}$		  \\ 
\noalign{\smallskip}
\noalign{\smallskip} 
$\gamma_{\mathrm{VIS}}$ [$\mathrm{m\,s^{-1}}$]          & 1.28$\,\pm\,$0.57 	    	& $1.25^{+0.37}_{-0.38}$ 		  \\ 
$\gamma_{\mathrm{HARPS}}$ [$\mathrm{m\,s^{-1}}$]        & 4.74$\,\pm\,$1.80 	    	& $4.96^{+1.63}_{-1.46}$ 		  \\ 
$\sigma_{\mathrm{jitt,VIS}}$ [$\mathrm{m\,s^{-1}}$]          & 2.4\ (fixed) 	    	& $2.40^{+0.37}_{-0.35}$ 		  \\ 
$\sigma_{\mathrm{jitt,HARPS}}$ [$\mathrm{m\,s^{-1}}$]        & 3.2\ (fixed) 	    	& $3.23^{+2.39}_{-2.10}$ 		  \\ 
\noalign{\smallskip}
\noalign{\smallskip}
$m_p\sin{i}$ [M$_{\oplus}$] & 7.4$\,\pm\,$0.5     		            & 		  \\ 
$a$ [au]                    & 0.026$\,\pm\,$0.001  	        	    & 		  \\ 
\noalign{\smallskip}
\hline
\end{tabular}
}
\end{table}

\begin{figure*}
\centering
\includegraphics[width=0.98\hsize, height=0.3\textheight]{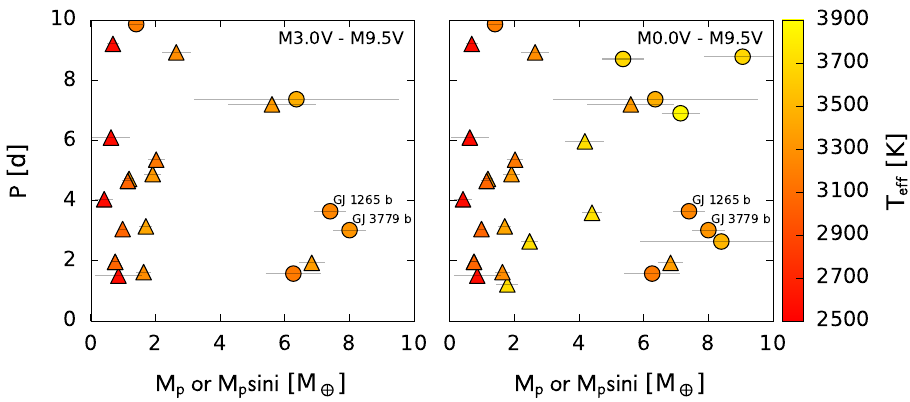}
\caption{The mass period diagram of all super-Earth planets, with masses determined from RVs or TTVs, orbiting mid- and late-type M dwarfs ({\it left}) and for all M dwarfs ({\it right}) taken from the NASA Exoplanet Archive as May 2018 (\texttt{https://exoplanetarchive.ipac.caltech.edu}). Single- and multi-planet systems are drawn with circles and triangles, respectively. The colorbar indicates the effective temperature of the host star.
} \label{fig:comparison}
\end{figure*}

The results from the radial velocity analysis reveal two very similar planets orbiting very similar M dwarfs. Not only do the stars exhibit comparable photospheric and physical parameters, but also long rotational periods. Mid-type slow-rotators are considered to be less magnetically active than their late-type or rapid-rotator counterparts \citep{2015ApJ...812....3W}, which agrees with the absence of periodic signals in the photospheric spectral indicators (Figs.~\ref{fig:gj3779_gls} and \ref{fig:gj1265_gls}). The planetary candidates are both orbiting at a semi-major axes of 0.026\,au with periods of the order of 3\,d and super-Earth-like minimum masses of 7--8\,M$_{\oplus}$. The MCMC analyses reveal that the derived eccentricities of GJ~3779~b and GJ~1265~b are compatible with circular orbits, which is predicted by orbital evolution models for such short-period planets.

Mass determination for Earth-size and super-Earth planets around M dwarfs is very challenging, and there is a limited number of systems similar to the ones presented here. Only eighteen planetary systems -- 
GJ~54.1 \citep{YZCet}, 
GJ~1132 \citep{GJ1132}, 
GJ~1214 \citep{GJ1214}, 
GJ~176 \citep{GJ176}, 
GJ~273 \citep{GJ273}, 
GJ~3138 \citep{GJ273}, 
GJ~3323 \citep{GJ273}, 
GJ~3634 \citep{GJ3634}, 
GJ~3942 \citep{GJ3942}, 
GJ~3998 \citep{GJ3998}, 
GJ~433 \citep{Bonfils13}, 
GJ~447 \citep{Ross128}, 
GJ~536 \citep{GJ536}, 
GJ~581 \citep{GJ581}, 
GJ~628 \citep{Wolf1061}, 
GJ~676~A \citep{GJ676A}, 
GJ~667~C \citep{Bonfils13}, 
GJ~876 \citep{Rivera05} -- have masses derived from radial velocity measurements. Furthermore, the TRAPPIST-1 system was detected by transit search and the planet's dynamical masses have been inferred from transit timing variations \citep{TRAPPIST1,TRAPPIST_MASS}. 

The mass-period parameter space of all known M dwarf planets with both parameters well-determined is shown in Fig.~\ref{fig:comparison}. The left panel shows the planets orbiting mid- and late-type M stars, while the right panel also includes those around early-type M dwarfs. The diagrams exhibit a bimodal distribution that resembles the one found in planetary radius using {\it Kepler} data  mainly for solar-like stars \citep{Fulton17,Fulton18}, but also confirmed for validated and well-characterized transiting planets around M dwarfs \citep{2018AJ....155..127H}.

It is also noticeable how short-period Earth-like planets in the range of 0.5--2\,M$_{\oplus}$ have been mostly found in multiple systems (triangle symbols), while in the range of 5--8\,M$_{\oplus}$, eight out of ten planets have not been found to have further planetary companions. Also striking is the lack of planets in the range of 2--5\,M$_{\oplus}$ with orbital periods shorter than 10\,d around mid- and late-type M stars. This raises questions about their formation process. Is it possible that super-Earth planets around M dwarfs prevent the formation of smaller counterparts, or are they formed by aggregation of smaller, Earth-size planets? Is it dependent on the mass of the host star?

The sample of known planets with accurate mass and radius determinations around M dwarfs is still too small to address with statistical confidence the validity of such distribution and its connection with planet multiplicity. Ground-based searches such as the CARMENES GTO program among others, together with planet detections around M dwarfs over the next few years from {\it TESS} \citep{TESS}, may provide insight into this issue.

In summary, we report on the discovery of two short-period super-Earth like planets orbiting the low mass stars GJ~3779 and GJ~1265. They are the least massive planets discovered by the CARMENES search for exoplanets around M dwarfs to date. The CARMENES visible Doppler measurements reveal a planetary companion with a minimum mass of 8.0\,M$_{\oplus}$ on a 3.023\,d orbit around GJ~3779, and a planet of 7.4\,M$_{\oplus}$ on a 3.651\,d orbit around GJ~1265. Aside the periodic signals at around 95\,d associated to rotation, the host stars do not show any evidence of RV-induced variations due to activity at the planet periods. The planets do not transit their parent stars, and they seem to belong to a population of short-period single planets around M stars with masses between 5--8\,M$_{\oplus}$ .

\begin{acknowledgements}
CARMENES is an instrument for the Centro Astron\'omico Hispano-Alem\'an de Calar Alto (CAHA, Almer\'{\i}a, Spain). CARMENES is funded by the German Max-Planck-Gesellschaft (MPG), the Spanish Consejo Superior de Investigaciones Cient\'{\i}ficas (CSIC), the European Union through FEDER/ERF FICTS-2011-02 funds, and the members of the CARMENES Consortium (Max-Planck-Institut f\"ur Astronomie, Instituto de Astrof\'{\i}sica de Andaluc\'{\i}a, Landessternwarte K\"onigstuhl, Institut de Ci\`encies de l'Espai, Insitut f\"ur Astrophysik G\"ottingen, Universidad Complutense de Madrid, Th\"uringer Landessternwarte Tautenburg, Instituto de Astrof\'{\i}sica de Canarias, Hamburger Sternwarte, Centro de Astrobiolog\'{\i}a and Centro Astron\'omico Hispano-Alem\'an), with additional contributions by the Spanish Ministry of Economy, the German Science Foundation through the Major Research Instrumentation Programme and DFG Research Unit FOR2544 ``Blue Planets around Red Stars'', the Klaus Tschira Stiftung, the states of Baden-W\"urttemberg and Niedersachsen, and by the Junta de Andaluc\'{\i}a. Based on observations collected at the European Organisation for Astronomical Research in the Southern Hemisphere under ESO programme(s) 072.C-0488(E) and 183.C-0437(A). R.L. has received funding from the European Union’s Horizon 2020 research and innovation program under the Marie Skłodowska-Curie grant agreement No. 713673 and financial support through the “la Caixa” INPhINIT Fellowship Grant for Doctoral studies at Spanish Research Centres of Excellence, “la Caixa” Banking Foundation, Barcelona, Spain. This work is partly financed by the Spanish Ministry of Economy and Competitiveness through grants ESP2013-48391-C4-2-R, ESP2016-80435-C2-1/2-R, and AYA2016-79425–C3–1/2/3–P.
\end{acknowledgements}

\bibliographystyle{aa} 
\bibliography{biblio} 

\begin{thebibliography}{69}
\expandafter\ifx\csname natexlab\endcsname\relax\def\natexlab#1{#1}\fi

\bibitem[{{Affer} {et~al.}(2016){Affer}, {Micela}, {Damasso}, {Perger},
  {Ribas}, {Su{\'a}rez Mascare{\~n}o}, {Gonz{\'a}lez Hern{\'a}ndez}, {Rebolo},
  {Poretti}, {Maldonado}, {Leto}, {Pagano}, {Scandariato}, {Zanmar Sanchez},
  {Sozzetti}, {Bonomo}, {Malavolta}, {Morales}, {Rosich}, {Bignamini},
  {Gratton}, {Velasco}, {Cenadelli}, {Claudi}, {Cosentino}, {Desidera},
  {Giacobbe}, {Herrero}, {Lafarga}, {Lanza}, {Molinari}, \& {Piotto}}]{GJ3998}
{Affer}, L., {Micela}, G., {Damasso}, M., {et~al.} 2016, \aap, 593, A117

\bibitem[{{Anglada-Escud{\'e}} {et~al.}(2016){Anglada-Escud{\'e}}, {Amado},
  {Barnes}, {Berdi{\~n}as}, {Butler}, {Coleman}, {de La Cueva}, {Dreizler},
  {Endl}, {Giesers}, {Jeffers}, {Jenkins}, {Jones}, {Kiraga}, {K{\"u}rster},
  {L{\'o}pez-Gonz{\'a}lez}, {Marvin}, {Morales}, {Morin}, {Nelson}, {Ortiz},
  {Ofir}, {Paardekooper}, {Reiners}, {Rodr{\'{\i}}guez},
  {Rodr{\'{\i}}guez-L{\'o}pez}, {Sarmiento}, {Strachan}, {Tsapras}, {Tuomi}, \&
  {Zechmeister}}]{ProxCen}
{Anglada-Escud{\'e}}, G., {Amado}, P.~J., {Barnes}, J., {et~al.} 2016, \nat,
  536, 437

\bibitem[{{Anglada-Escud{\'e}} \& {Tuomi}(2012)}]{GJ676A}
{Anglada-Escud{\'e}}, G. \& {Tuomi}, M. 2012, \aap, 548, A58

\bibitem[{{Astudillo-Defru} {et~al.}(2017{\natexlab{a}}){Astudillo-Defru},
  {D{\'{\i}}az}, {Bonfils}, {Almenara}, {Delisle}, {Bouchy}, {Delfosse},
  {Forveille}, {Lovis}, {Mayor}, {Murgas}, {Pepe}, {Santos}, {S{\'e}gransan},
  {Udry}, \& {W{\"u}nsche}}]{YZCet}
{Astudillo-Defru}, N., {D{\'{\i}}az}, R.~F., {Bonfils}, X., {et~al.}
  2017{\natexlab{a}}, \aap, 605, L11

\bibitem[{{Astudillo-Defru} {et~al.}(2017{\natexlab{b}}){Astudillo-Defru},
  {Forveille}, {Bonfils}, {S{\'e}gransan}, {Bouchy}, {Delfosse}, {Lovis},
  {Mayor}, {Murgas}, {Pepe}, {Santos}, {Udry}, \& {W{\"u}nsche}}]{GJ273}
{Astudillo-Defru}, N., {Forveille}, T., {Bonfils}, X., {et~al.}
  2017{\natexlab{b}}, \aap, 602, A88

\bibitem[{{Baluev}(2009)}]{Baluev09}
{Baluev}, R.~V. 2009, \mnras, 393, 969

\bibitem[{{Barnes} {et~al.}(2011){Barnes}, {Jeffers}, \&
  {Jones}}]{2011MNRAS.412.1599B}
{Barnes}, J.~R., {Jeffers}, S.~V., \& {Jones}, H.~R.~A. 2011, \mnras, 412, 1599

\bibitem[{{Bayo} {et~al.}(2008){Bayo}, {Rodrigo}, {Barrado Y Navascu{\'e}s},
  {Solano}, {Guti{\'e}rrez}, {Morales-Calder{\'o}n}, \& {Allard}}]{Bayo08}
{Bayo}, A., {Rodrigo}, C., {Barrado Y Navascu{\'e}s}, D., {et~al.} 2008, \aap,
  492, 277

\bibitem[{{Benedict} {et~al.}(2016){Benedict}, {Henry}, {Franz}, {McArthur},
  {Wasserman}, {Jao}, {Cargile}, {Dieterich}, {Bradley}, {Nelan}, \&
  {Whipple}}]{Benedict16}
{Benedict}, G.~F., {Henry}, T.~J., {Franz}, O.~G., {et~al.} 2016, \aj, 152, 141

\bibitem[{{Berta} {et~al.}(2012){Berta}, {Irwin}, {Charbonneau}, {Burke}, \&
  {Falco}}]{MEarth12}
{Berta}, Z.~K., {Irwin}, J., {Charbonneau}, D., {Burke}, C.~J., \& {Falco},
  E.~E. 2012, \aj, 144, 145

\bibitem[{{Berta-Thompson} {et~al.}(2015){Berta-Thompson}, {Irwin},
  {Charbonneau}, {Newton}, {Dittmann}, {Astudillo-Defru}, {Bonfils}, {Gillon},
  {Jehin}, {Stark}, {Stalder}, {Bouchy}, {Delfosse}, {Forveille}, {Lovis},
  {Mayor}, {Neves}, {Pepe}, {Santos}, {Udry}, \& {W{\"u}nsche}}]{GJ1132}
{Berta-Thompson}, Z.~K., {Irwin}, J., {Charbonneau}, D., {et~al.} 2015, \nat,
  527, 204

\bibitem[{{Bonfils} {et~al.}(2018){Bonfils}, {Astudillo-Defru}, {D{\'{\i}}az},
  {Almenara}, {Forveille}, {Bouchy}, {Delfosse}, {Lovis}, {Mayor}, {Murgas},
  {Pepe}, {Santos}, {S{\'e}gransan}, {Udry}, \& {W{\"u}nsche}}]{Ross128}
{Bonfils}, X., {Astudillo-Defru}, N., {D{\'{\i}}az}, R., {et~al.} 2018, \aap,
  613, A25

\bibitem[{{Bonfils} {et~al.}(2013){Bonfils}, {Delfosse}, {Udry}, {Forveille},
  {Mayor}, {Perrier}, {Bouchy}, {Gillon}, {Lovis}, {Pepe}, {Queloz}, {Santos},
  {S{\'e}gransan}, \& {Bertaux}}]{Bonfils13}
{Bonfils}, X., {Delfosse}, X., {Udry}, S., {et~al.} 2013, \aap, 549, A109

\bibitem[{{Bonfils} {et~al.}(2005){Bonfils}, {Forveille}, {Delfosse}, {Udry},
  {Mayor}, {Perrier}, {Bouchy}, {Pepe}, {Queloz}, \& {Bertaux}}]{GJ581}
{Bonfils}, X., {Forveille}, T., {Delfosse}, X., {et~al.} 2005, \aap, 443, L15

\bibitem[{{Bonfils} {et~al.}(2011){Bonfils}, {Gillon}, {Forveille}, {Delfosse},
  {Deming}, {Demory}, {Lovis}, {Mayor}, {Neves}, {Perrier}, {Santos}, {Seager},
  {Udry}, {Boisse}, \& {Bonnefoy}}]{GJ3634}
{Bonfils}, X., {Gillon}, M., {Forveille}, T., {et~al.} 2011, \aap, 528, A111

\bibitem[{{Charbonneau} {et~al.}(2009){Charbonneau}, {Berta}, {Irwin}, {Burke},
  {Nutzman}, {Buchhave}, {Lovis}, {Bonfils}, {Latham}, {Udry}, {Murray-Clay},
  {Holman}, {Falco}, {Winn}, {Queloz}, {Pepe}, {Mayor}, {Delfosse}, \&
  {Forveille}}]{GJ1214}
{Charbonneau}, D., {Berta}, Z.~K., {Irwin}, J., {et~al.} 2009, \nat, 462, 891

\bibitem[{{Charbonneau} {et~al.}(2008){Charbonneau}, {Irwin}, {Nutzman}, \&
  {Falco}}]{MEarth}
{Charbonneau}, D., {Irwin}, J., {Nutzman}, P., \& {Falco}, E.~E. 2008, in
  Bulletin of the American Astronomical Society, Vol.~40, American Astronomical
  Society Meeting Abstracts \#212, 242

\bibitem[{{Cortés-Contreras}(2016)}]{CC16}
{Cortés-Contreras}, M. 2016, PhD thesis, Universidad Complutense de Madrid,
  Spain

\bibitem[{{Delfosse} {et~al.}(2000){Delfosse}, {Forveille}, {S{\'e}gransan},
  {Beuzit}, {Udry}, {Perrier}, \& {Mayor}}]{Delfosse00}
{Delfosse}, X., {Forveille}, T., {S{\'e}gransan}, D., {et~al.} 2000, \aap, 364,
  217

\bibitem[{{Dressing} \& {Charbonneau}(2015)}]{Dress15}
{Dressing}, C.~D. \& {Charbonneau}, D. 2015, \apj, 807, 45

\bibitem[{{Foreman-Mackey} {et~al.}(2013){Foreman-Mackey}, {Hogg}, {Lang}, \&
  {Goodman}}]{emcee}
{Foreman-Mackey}, D., {Hogg}, D.~W., {Lang}, D., \& {Goodman}, J. 2013, \pasp,
  125, 306

\bibitem[{{Forveille} {et~al.}(2009){Forveille}, {Bonfils}, {Delfosse},
  {Gillon}, {Udry}, {Bouchy}, {Lovis}, {Mayor}, {Pepe}, {Perrier}, {Queloz},
  {Santos}, \& {Bertaux}}]{GJ176}
{Forveille}, T., {Bonfils}, X., {Delfosse}, X., {et~al.} 2009, \aap, 493, 645

\bibitem[{{Fulton} \& {Petigura}(2018)}]{Fulton18}
{Fulton}, B.~J. \& {Petigura}, E.~A. 2018, ArXiv e-prints
  [\eprint[arXiv]{1805.01453}]

\bibitem[{{Fulton} {et~al.}(2017){Fulton}, {Petigura}, {Howard}, {Isaacson},
  {Marcy}, {Cargile}, {Hebb}, {Weiss}, {Johnson}, {Morton}, {Sinukoff},
  {Crossfield}, \& {Hirsch}}]{Fulton17}
{Fulton}, B.~J., {Petigura}, E.~A., {Howard}, A.~W., {et~al.} 2017, \aj, 154,
  109

\bibitem[{{Gaia Collaboration} {et~al.}(2018){Gaia Collaboration}, {Brown},
  {Vallenari}, {Prusti}, {de Bruijne}, {Babusiaux}, {Bailer-Jones}, {Biermann},
  {Evans}, {Eyer}, \& et~al.}]{GaiaDR2}
{Gaia Collaboration}, {Brown}, A.~G.~A., {Vallenari}, A., {et~al.} 2018, \aap,
  616, A1

\bibitem[{{Gaidos} {et~al.}(2016){Gaidos}, {Mann}, {Kraus}, \&
  {Ireland}}]{2016MNRAS.457.2877G}
{Gaidos}, E., {Mann}, A.~W., {Kraus}, A.~L., \& {Ireland}, M. 2016, \mnras,
  457, 2877

\bibitem[{{Gillon} {et~al.}(2017){Gillon}, {Triaud}, {Demory}, {Jehin}, {Agol},
  {Deck}, {Lederer}, {de Wit}, {Burdanov}, {Ingalls}, {Bolmont}, {Leconte},
  {Raymond}, {Selsis}, {Turbet}, {Barkaoui}, {Burgasser}, {Burleigh}, {Carey},
  {Chaushev}, {Copperwheat}, {Delrez}, {Fernandes}, {Holdsworth}, {Kotze}, {Van
  Grootel}, {Almleaky}, {Benkhaldoun}, {Magain}, \& {Queloz}}]{TRAPPIST1}
{Gillon}, M., {Triaud}, A.~H.~M.~J., {Demory}, B.-O., {et~al.} 2017, \nat, 542,
  456

\bibitem[{{Grimm} {et~al.}(2018){Grimm}, {Demory}, {Gillon}, {Dorn}, {Agol},
  {Burdanov}, {Delrez}, {Sestovic}, {Triaud}, {Turbet}, {Bolmont}, {Caldas},
  {Wit}, {Jehin}, {Leconte}, {Raymond}, {Grootel}, {Burgasser}, {Carey},
  {Fabrycky}, {Heng}, {Hernandez}, {Ingalls}, {Lederer}, {Selsis}, \&
  {Queloz}}]{TRAPPIST_MASS}
{Grimm}, S.~L., {Demory}, B.-O., {Gillon}, M., {et~al.} 2018, \aap, 613, A68

\bibitem[{{Hawley} {et~al.}(1996){Hawley}, {Gizis}, \& {Reid}}]{PMSU}
{Hawley}, S.~L., {Gizis}, J.~E., \& {Reid}, I.~N. 1996, \aj, 112, 2799

\bibitem[{{Henry} {et~al.}(2016){Henry}, {Jao}, {Winters}, {Dieterich},
  {Finch}, {Hambly}, {Ianna}, {McCarthy}, {Riedel}, {Subasavage}, \& {RECONS
  Team}}]{2016AAS...22714201H}
{Henry}, T.~J., {Jao}, W.-C., {Winters}, J.~G., {et~al.} 2016, in American
  Astronomical Society Meeting Abstracts, Vol. 227, American Astronomical
  Society Meeting Abstracts \#227, 142.01

\bibitem[{{Hirano} {et~al.}(2018){Hirano}, {Dai}, {Gandolfi}, {Fukui},
  {Livingston}, {Miyakawa}, {Endl}, {Cochran}, {Alonso-Floriano}, {Kuzuhara},
  {Montes}, {Ryu}, {Albrecht}, {Barragan}, {Cabrera}, {Csizmadia}, {Deeg},
  {Eigm{\"u}ller}, {Erikson}, {Fridlund}, {Grziwa}, {Guenther}, {Hatzes},
  {Korth}, {Kudo}, {Kusakabe}, {Narita}, {Nespral}, {Nowak}, {P{\"a}tzold},
  {Palle}, {Persson}, {Prieto-Arranz}, {Rauer}, {Ribas}, {Sato}, {Smith},
  {Tamura}, {Tanaka}, {Van Eylen}, \& {Winn}}]{2018AJ....155..127H}
{Hirano}, T., {Dai}, F., {Gandolfi}, D., {et~al.} 2018, \aj, 155, 127

\bibitem[{{Husser} {et~al.}(2013){Husser}, {Wende-von Berg}, {Dreizler},
  {Homeier}, {Reiners}, {Barman}, \& {Hauschildt}}]{PHOENIX-ACES}
{Husser}, T.-O., {Wende-von Berg}, S., {Dreizler}, S., {et~al.} 2013, \aap,
  553, A6

\bibitem[{{Jeffers} {et~al.}(2018){Jeffers}, {Sch{\"o}fer}, {Lamert},
  {Reiners}, {Montes}, {Caballero}, {Cort{\'e}s-Contreras}, {Marvin},
  {Passegger}, {Zechmeister}, {Quirrenbach}, {Alonso-Floriano}, {Amado},
  {Bauer}, {Casal}, {Alonso}, {Herrero}, {Morales}, {Mundt}, {Ribas}, \&
  {Sarmiento}}]{Jeffers18}
{Jeffers}, S.~V., {Sch{\"o}fer}, P., {Lamert}, A., {et~al.} 2018, \aap, 614,
  A76

\bibitem[{{Kaminski} {et~al.}(2018){Kaminski}, {Trifonov}, {Caballero},
  {Quirrenbach}, {Ribas}, {Reiners}, {Amado}, {Zechmeister}, {Dreizler},
  {Perger}, {Tal-Or}, {Bonfils}, {Mayor}, {Astudillo-Defru}, {Bauer},
  {B{\'e}jar}, {Cifuentes}, {Colom{\'e}}, {Cort{\'e}s-Contreras}, {Delfosse},
  {D{\'{\i}}ez-Alonso}, {Forveille}, {Guenther}, {Hatzes}, {Henning},
  {Jeffers}, {K{\"u}rster}, {Lafarga}, {Luque}, {Mandel}, {Montes}, {Morales},
  {Passegger}, {Pedraz}, {Reffert}, {Sadegi}, {Schweitzer}, {Seifert}, {Stahl},
  \& {Udry}}]{Kaminski18}
{Kaminski}, A., {Trifonov}, T., {Caballero}, J.~A., {et~al.} 2018, ArXiv
  e-prints [\eprint[arXiv]{1808.01183}]

\bibitem[{{Kane} {et~al.}(2009){Kane}, {Mahadevan}, {von Braun}, {Laughlin}, \&
  {Ciardi}}]{Kane09}
{Kane}, S.~R., {Mahadevan}, S., {von Braun}, K., {Laughlin}, G., \& {Ciardi},
  D.~R. 2009, \pasp, 121, 1386

\bibitem[{{Kopparapu} {et~al.}(2013){Kopparapu}, {Ramirez}, {Kasting}, {Eymet},
  {Robinson}, {Mahadevan}, {Terrien}, {Domagal-Goldman}, {Meadows}, \&
  {Deshpande}}]{Kopparapu13}
{Kopparapu}, R.~K., {Ramirez}, R., {Kasting}, J.~F., {et~al.} 2013, \apj, 765,
  131

\bibitem[{{Kopparapu} {et~al.}(2014){Kopparapu}, {Ramirez}, {SchottelKotte},
  {Kasting}, {Domagal-Goldman}, \& {Eymet}}]{Kopparapu14}
{Kopparapu}, R.~K., {Ramirez}, R.~M., {SchottelKotte}, J., {et~al.} 2014,
  \apjl, 787, L29

\bibitem[{{Kov{\'a}cs} {et~al.}(2002){Kov{\'a}cs}, {Zucker}, \& {Mazeh}}]{BLS}
{Kov{\'a}cs}, G., {Zucker}, S., \& {Mazeh}, T. 2002, \aap, 391, 369

\bibitem[{{Marcy} {et~al.}(1998){Marcy}, {Butler}, {Vogt}, {Fischer}, \&
  {Lissauer}}]{Marcy98}
{Marcy}, G.~W., {Butler}, R.~P., {Vogt}, S.~S., {Fischer}, D., \& {Lissauer},
  J.~J. 1998, \apjl, 505, L147

\bibitem[{{Mayor} {et~al.}(2003){Mayor}, {Pepe}, {Queloz}, {Bouchy},
  {Rupprecht}, {Lo Curto}, {Avila}, {Benz}, {Bertaux}, {Bonfils}, {Dall},
  {Dekker}, {Delabre}, {Eckert}, {Fleury}, {Gilliotte}, {Gojak}, {Guzman},
  {Kohler}, {Lizon}, {Longinotti}, {Lovis}, {Megevand}, {Pasquini}, {Reyes},
  {Sivan}, {Sosnowska}, {Soto}, {Udry}, {van Kesteren}, {Weber}, \&
  {Weilenmann}}]{HARPS}
{Mayor}, M., {Pepe}, F., {Queloz}, D., {et~al.} 2003, The Messenger, 114, 20

\bibitem[{{Neves} {et~al.}(2014){Neves}, {Bonfils}, {Santos}, {Delfosse},
  {Forveille}, {Allard}, \& {Udry}}]{Neves14}
{Neves}, V., {Bonfils}, X., {Santos}, N.~C., {et~al.} 2014, \aap, 568, A121

\bibitem[{{Newton} {et~al.}(2014){Newton}, {Charbonneau}, {Irwin},
  {Berta-Thompson}, {Rojas-Ayala}, {Covey}, \& {Lloyd}}]{Newton14}
{Newton}, E.~R., {Charbonneau}, D., {Irwin}, J., {et~al.} 2014, \aj, 147, 20

\bibitem[{{Passegger} {et~al.}(2018){Passegger}, {Reiners}, {Jeffers},
  {Wende-von Berg}, {Sch{\"o}fer}, {Caballero}, {Schweitzer}, {Amado},
  {B{\'e}jar}, {Cort{\'e}s-Contreras}, {Hatzes}, {K{\"u}rster}, {Montes},
  {Pedraz}, {Quirrenbach}, {Ribas}, \& {Seifert}}]{Passegger18}
{Passegger}, V.~M., {Reiners}, A., {Jeffers}, S.~V., {et~al.} 2018, \aap, 615,
  A6

\bibitem[{{Passegger} {et~al.}(2016){Passegger}, {Wende-von Berg}, \&
  {Reiners}}]{Passegger16}
{Passegger}, V.~M., {Wende-von Berg}, S., \& {Reiners}, A. 2016, \aap, 587, A19

\bibitem[{{Perger} {et~al.}(2017){Perger}, {Ribas}, {Damasso}, {Morales},
  {Affer}, {Su{\'a}rez Mascare{\~n}o}, {Micela}, {Maldonado}, {Gonz{\'a}lez
  Hern{\'a}ndez}, {Rebolo}, {Scandariato}, {Leto}, {Zanmar Sanchez}, {Benatti},
  {Bignamini}, {Borsa}, {Carbognani}, {Claudi}, {Desidera}, {Esposito},
  {Lafarga}, {Martinez Fiorenzano}, {Herrero}, {Molinari}, {Nascimbeni},
  {Pagano}, {Pedani}, {Poretti}, {Rainer}, {Rosich}, {Sozzetti}, \&
  {Toledo-Padr{\'o}n}}]{GJ3942}
{Perger}, M., {Ribas}, I., {Damasso}, M., {et~al.} 2017, \aap, 608, A63

\bibitem[{{Press} {et~al.}(1992){Press}, {Teukolsky}, {Vetterling}, \&
  {Flannery}}]{1992nrfa.book.....P}
{Press}, W.~H., {Teukolsky}, S.~A., {Vetterling}, W.~T., \& {Flannery}, B.~P.
  1992, {Numerical recipes in FORTRAN. The art of scientific computing}

\bibitem[{{Queloz} {et~al.}(2001){Queloz}, {Henry}, {Sivan}, {Baliunas},
  {Beuzit}, {Donahue}, {Mayor}, {Naef}, {Perrier}, \&
  {Udry}}]{2001A&A...379..279Q}
{Queloz}, D., {Henry}, G.~W., {Sivan}, J.~P., {et~al.} 2001, \aap, 379, 279

\bibitem[{{Quirrenbach} {et~al.}(2014){Quirrenbach}, {Amado}, {Caballero},
  {Mundt}, {Reiners}, {Ribas}, {Seifert}, {Abril}, {Aceituno},
  {Alonso-Floriano}, {Ammler-von Eiff}, {Antona Jim{\'e}nez},
  {Anwand-Heerwart}, {Azzaro}, {Bauer}, {Barrado}, {Becerril}, {B{\'e}jar},
  {Ben{\'{\i}}tez}, {Berdi{\~n}as}, {C{\'a}rdenas}, {Casal}, {Claret},
  {Colom{\'e}}, {Cort{\'e}s-Contreras}, {Czesla}, {Doellinger}, {Dreizler},
  {Feiz}, {Fern{\'a}ndez}, {Galad{\'{\i}}}, {G{\'a}lvez-Ortiz},
  {Garc{\'{\i}}a-Piquer}, {Garc{\'{\i}}a-Vargas}, {Garrido}, {Gesa}, {G{\'o}mez
  Galera}, {Gonz{\'a}lez {\'A}lvarez}, {Gonz{\'a}lez Hern{\'a}ndez},
  {Gr{\"o}zinger}, {Gu{\`a}rdia}, {Guenther}, {de Guindos},
  {Guti{\'e}rrez-Soto}, {Hagen}, {Hatzes}, {Hauschildt}, {Helmling}, {Henning},
  {Hermann}, {Hern{\'a}ndez Casta{\~n}o}, {Herrero}, {Hidalgo}, {Holgado},
  {Huber}, {Huber}, {Jeffers}, {Joergens}, {de Juan}, {Kehr}, {Klein},
  {K{\"u}rster}, {Lamert}, {Lalitha}, {Laun}, {Lemke}, {Lenzen}, {L{\'o}pez del
  Fresno}, {L{\'o}pez Mart{\'{\i}}}, {L{\'o}pez-Santiago}, {Mall}, {Mandel},
  {Mart{\'{\i}}n}, {Mart{\'{\i}}n-Ruiz}, {Mart{\'{\i}}nez-Rodr{\'{\i}}guez},
  {Marvin}, {Mathar}, {Mirabet}, {Montes}, {Morales Mu{\~n}oz}, {Moya},
  {Naranjo}, {Ofir}, {Oreiro}, {Pall{\'e}}, {Panduro}, {Passegger},
  {P{\'e}rez-Calpena}, {P{\'e}rez Medialdea}, {Perger}, {Pluto}, {Ram{\'o}n},
  {Rebolo}, {Redondo}, {Reffert}, {Reinhardt}, {Rhode}, {Rix}, {Rodler},
  {Rodr{\'{\i}}guez}, {Rodr{\'{\i}}guez-L{\'o}pez},
  {Rodr{\'{\i}}guez-P{\'e}rez}, {Rohloff}, {Rosich}, {S{\'a}nchez-Blanco},
  {S{\'a}nchez Carrasco}, {Sanz-Forcada}, {Sarmiento}, {Sch{\"a}fer},
  {Schiller}, {Schmidt}, {Schmitt}, {Solano}, {Stahl}, {Storz}, {St{\"u}rmer},
  {Su{\'a}rez}, {Ulbrich}, {Veredas}, {Wagner}, {Winkler}, {Zapatero Osorio},
  {Zechmeister}, {Abell{\'a}n de Paco}, {Anglada-Escud{\'e}}, {del Burgo},
  {Klutsch}, {Lizon}, {L{\'o}pez-Morales}, {Morales}, {Perryman}, {Tulloch}, \&
  {Xu}}]{CARMENES}
{Quirrenbach}, A., {Amado}, P.~J., {Caballero}, J.~A., {et~al.} 2014, in
  \procspie, Vol. 9147, Ground-based and Airborne Instrumentation for Astronomy
  V, 91471F

\bibitem[{{Reiners} {et~al.}(2018{\natexlab{a}}){Reiners}, {Ribas},
  {Zechmeister}, {Caballero}, {Trifonov}, {Dreizler}, {Morales}, {Tal-Or},
  {Lafarga}, {Quirrenbach}, {Amado}, {Kaminski}, {Jeffers}, {Aceituno},
  {B{\'e}jar}, {Gu{\`a}rdia}, {Guenther}, {Hagen}, {Montes}, {Passegger},
  {Seifert}, {Schweitzer}, {Cort{\'e}s-Contreras}, {Abril}, {Alonso-Floriano},
  {Eiff}, {Antona}, {Anglada-Escud{\'e}}, {Anwand-Heerwart}, {Arroyo-Torres},
  {Azzaro}, {Baroch}, {Barrado}, {Bauer}, {Becerril}, {Ben{\'{\i}}tez},
  {Berdi{\~n}as}, {Bergond}, {Bl{\"u}mcke}, {Brinkm{\"o}ller}, {del Burgo},
  {Cano}, {C{\'a}rdenas V{\'a}zquez}, {Casal}, {Cifuentes}, {Claret},
  {Colom{\'e}}, {Czesla}, {D{\'{\i}}ez-Alonso}, {Feiz}, {Fern{\'a}ndez},
  {Ferro}, {Fuhrmeister}, {Galad{\'{\i}}-Enr{\'{\i}}quez}, {Garcia-Piquer},
  {Garc{\'{\i}}a Vargas}, {Gesa}, {G{\'o}mez Galera}, {Gonz{\'a}lez
  Hern{\'a}ndez}, {Gonz{\'a}lez-Peinado}, {Gr{\"o}zinger}, {Grohnert},
  {Guijarro}, {de Guindos}, {Guti{\'e}rrez-Soto}, {Hatzes}, {Hauschildt},
  {Hedrosa}, {Helmling}, {Henning}, {Hermelo}, {Hern{\'a}ndez Arab{\'{\i}}},
  {Hern{\'a}ndez Casta{\~n}o}, {Hern{\'a}ndez Hernando}, {Herrero}, {Huber},
  {Huke}, {Johnson}, {de Juan}, {Kim}, {Klein}, {Kl{\"u}ter}, {Klutsch},
  {K{\"u}rster}, {Labarga}, {Lamert}, {Lamp{\'o}n}, {Lara}, {Laun}, {Lemke},
  {Lenzen}, {Launhardt}, {L{\'o}pez del Fresno}, {L{\'o}pez-Gonz{\'a}lez},
  {L{\'o}pez-Puertas}, {L{\'o}pez Salas}, {L{\'o}pez-Santiago}, {Luque},
  {Mag{\'a}n Madinabeitia}, {Mall}, {Mancini}, {Mandel}, {Marfil},
  {Mar{\'{\i}}n Molina}, {Maroto Fern{\'a}ndez}, {Mart{\'{\i}}n},
  {Mart{\'{\i}}n-Ruiz}, {Marvin}, {Mathar}, {Mirabet}, {Moreno-Raya}, {Moya},
  {Mundt}, {Nagel}, {Naranjo}, {Nortmann}, {Nowak}, {Ofir}, {Oreiro},
  {Pall{\'e}}, {Panduro}, {Pascual}, {Pavlov}, {Pedraz}, {P{\'e}rez-Calpena},
  {P{\'e}rez Medialdea}, {Perger}, {Perryman}, {Pluto}, {Rabaza}, {Ram{\'o}n},
  {Rebolo}, {Redondo}, {Reffert}, {Reinhart}, {Rhode}, {Rix}, {Rodler},
  {Rodr{\'{\i}}guez}, {Rodr{\'{\i}}guez-L{\'o}pez}, {Rodr{\'{\i}}guez
  Trinidad}, {Rohloff}, {Rosich}, {Sadegi}, {S{\'a}nchez-Blanco}, {S{\'a}nchez
  Carrasco}, {S{\'a}nchez-L{\'o}pez}, {Sanz-Forcada}, {Sarkis}, {Sarmiento},
  {Sch{\"a}fer}, {Schmitt}, {Schiller}, {Sch{\"o}fer}, {Solano}, {Stahl},
  {Strachan}, {St{\"u}rmer}, {Su{\'a}rez}, {Tabernero}, {Tala}, {Tulloch},
  {Ulbrich}, {Veredas}, {Vico Linares}, {Vilardell}, {Wagner}, {Winkler},
  {Wolthoff}, {Xu}, {Yan}, \& {Zapatero Osorio}}]{Reiners18}
{Reiners}, A., {Ribas}, I., {Zechmeister}, M., {et~al.} 2018{\natexlab{a}},
  \aap, 609, L5

\bibitem[{{Reiners} {et~al.}(2014){Reiners}, {Sch{\"u}ssler}, \&
  {Passegger}}]{Reiners14}
{Reiners}, A., {Sch{\"u}ssler}, M., \& {Passegger}, V.~M. 2014, \apj, 794, 144

\bibitem[{{Reiners} {et~al.}(2018{\natexlab{b}}){Reiners}, {Zechmeister},
  {Caballero}, {Ribas}, {Morales}, {Jeffers}, {Sch{\"o}fer}, {Tal-Or},
  {Quirrenbach}, {Amado}, {Kaminski}, {Seifert}, {Abril}, {Aceituno},
  {Alonso-Floriano}, {Ammler-von Eiff}, {Antona}, {Anglada-Escud{\'e}},
  {Anwand-Heerwart}, {Arroyo-Torres}, {Azzaro}, {Baroch}, {Barrado}, {Bauer},
  {Becerril}, {B{\'e}jar}, {Ben{\'{\i}}tez}, {Berdi{\~n}as}, {Bergond},
  {Bl{\"u}mcke}, {Brinkm{\"o}ller}, {del Burgo}, {Cano}, {C{\'a}rdenas
  V{\'a}zquez}, {Casal}, {Cifuentes}, {Claret}, {Colom{\'e}},
  {Cort{\'e}s-Contreras}, {Czesla}, {D{\'{\i}}ez-Alonso}, {Dreizler}, {Feiz},
  {Fern{\'a}ndez}, {Ferro}, {Fuhrmeister}, {Galad{\'{\i}}-Enr{\'{\i}}quez},
  {Garcia-Piquer}, {Garc{\'{\i}}a Vargas}, {Gesa}, {G{\'o}mez}, {Galera},
  {Gonz{\'a}lez Hern{\'a}ndez}, {Gonz{\'a}lez-Peinado}, {Gr{\"o}zinger},
  {Grohnert}, {Gu{\`a}rdia}, {Guenther}, {Guijarro}, {de Guindos},
  {Guti{\'e}rrez-Soto}, {Hagen}, {Hatzes}, {Hauschildt}, {Hedrosa}, {Helmling},
  {Henning}, {Hermelo}, {Hern{\'a}ndez Arab{\'{\i}}}, {Hern{\'a}ndez
  Casta{\~n}o}, {Hern{\'a}ndez Hernando}, {Herrero}, {Huber}, {Huke},
  {Johnson}, {de Juan}, {Kim}, {Klein}, {Kl{\"u}ter}, {Klutsch}, {K{\"u}rster},
  {Lafarga}, {Lamert}, {Lamp{\'o}n}, {Lara}, {Laun}, {Lemke}, {Lenzen},
  {Launhardt}, {L{\'o}pez del Fresno}, {L{\'o}pez-Gonz{\'a}lez},
  {L{\'o}pez-Puertas}, {L{\'o}pez Salas}, {L{\'o}pez-Santiago}, {Luque},
  {Mag{\'a}n Madinabeitia}, {Mall}, {Mancini}, {Mandel}, {Marfil},
  {Mar{\'{\i}}n Molina}, {Maroto}, {Fern{\'a}ndez}, {Mart{\'{\i}}n},
  {Mart{\'{\i}}n-Ruiz}, {Marvin}, {Mathar}, {Mirabet}, {Montes}, {Moreno-Raya},
  {Moya}, {Mundt}, {Nagel}, {Naranjo}, {Nortmann}, {Nowak}, {Ofir}, {Oreiro},
  {Pall{\'e}}, {Panduro}, {Pascual}, {Passegger}, {Pavlov}, {Pedraz},
  {P{\'e}rez-Calpena}, {P{\'e}rez Medialdea}, {Perger}, {Perryman}, {Pluto},
  {Rabaza}, {Ram{\'o}n}, {Rebolo}, {Redondo}, {Reffert}, {Reinhart}, {Rhode},
  {Rix}, {Rodler}, {Rodr{\'{\i}}guez}, {Rodr{\'{\i}}guez-L{\'o}pez},
  {Rodr{\'{\i}}guez Trinidad}, {Rohloff}, {Rosich}, {Sadegi},
  {S{\'a}nchez-Blanco}, {S{\'a}nchez Carrasco}, {S{\'a}nchez-L{\'o}pez},
  {Sanz-Forcada}, {Sarkis}, {Sarmiento}, {Sch{\"a}fer}, {Schmitt}, {Schiller},
  {Schweitzer}, {Solano}, {Stahl}, {Strachan}, {St{\"u}rmer}, {Su{\'a}rez},
  {Tabernero}, {Tala}, {Trifonov}, {Tulloch}, {Ulbrich}, {Veredas}, {Vico
  Linares}, {Vilardell}, {Wagner}, {Winkler}, {Wolthoff}, {Xu}, {Yan}, \&
  {Zapatero Osorio}}]{Reiners17}
{Reiners}, A., {Zechmeister}, M., {Caballero}, J.~A., {et~al.}
  2018{\natexlab{b}}, \aap, 612, A49

\bibitem[{{Ricker} {et~al.}(2014){Ricker}, {Winn}, {Vanderspek}, {Latham},
  {Bakos}, {Bean}, {Berta-Thompson}, {Brown}, {Buchhave}, {Butler}, {Butler},
  {Chaplin}, {Charbonneau}, {Christensen-Dalsgaard}, {Clampin}, {Deming},
  {Doty}, {De Lee}, {Dressing}, {Dunham}, {Endl}, {Fressin}, {Ge}, {Henning},
  {Holman}, {Howard}, {Ida}, {Jenkins}, {Jernigan}, {Johnson}, {Kaltenegger},
  {Kawai}, {Kjeldsen}, {Laughlin}, {Levine}, {Lin}, {Lissauer}, {MacQueen},
  {Marcy}, {McCullough}, {Morton}, {Narita}, {Paegert}, {Palle}, {Pepe},
  {Pepper}, {Quirrenbach}, {Rinehart}, {Sasselov}, {Sato}, {Seager},
  {Sozzetti}, {Stassun}, {Sullivan}, {Szentgyorgyi}, {Torres}, {Udry}, \&
  {Villasenor}}]{TESS}
{Ricker}, G.~R., {Winn}, J.~N., {Vanderspek}, R., {et~al.} 2014, in \procspie,
  Vol. 9143, Space Telescopes and Instrumentation 2014: Optical, Infrared, and
  Millimeter Wave, 914320

\bibitem[{{Rivera} {et~al.}(2005){Rivera}, {Lissauer}, {Butler}, {Marcy},
  {Vogt}, {Fischer}, {Brown}, {Laughlin}, \& {Henry}}]{Rivera05}
{Rivera}, E.~J., {Lissauer}, J.~J., {Butler}, R.~P., {et~al.} 2005, \apj, 634,
  625

\bibitem[{{Robertson} {et~al.}(2014){Robertson}, {Mahadevan}, {Endl}, \&
  {Roy}}]{2014Sci...345..440R}
{Robertson}, P., {Mahadevan}, S., {Endl}, M., \& {Roy}, A. 2014, Science, 345,
  440

\bibitem[{{Rosen} {et~al.}(2016){Rosen}, {Webb}, {Watson}, {Ballet}, {Barret},
  {Braito}, {Carrera}, {Ceballos}, {Coriat}, {Della Ceca}, {Denkinson},
  {Esquej}, {Farrell}, {Freyberg}, {Gris{\'e}}, {Guillout}, {Heil},
  {Koliopanos}, {Law-Green}, {Lamer}, {Lin}, {Martino}, {Michel}, {Motch},
  {Nebot Gomez-Moran}, {Page}, {Page}, {Page}, {Pakull}, {Pye}, {Read},
  {Rodriguez}, {Sakano}, {Saxton}, {Schwope}, {Scott}, {Sturm}, {Traulsen},
  {Yershov}, \& {Zolotukhin}}]{3XMM}
{Rosen}, S.~R., {Webb}, N.~A., {Watson}, M.~G., {et~al.} 2016, \aap, 590, A1

\bibitem[{{Sarkis} {et~al.}(2018){Sarkis}, {Henning}, {K{\"u}rster},
  {Trifonov}, {Zechmeister}, {Tal-Or}, {Anglada-Escud{\'e}}, {Hatzes},
  {Lafarga}, {Dreizler}, {Ribas}, {Caballero}, {Reiners}, {Mallonn}, {Morales},
  {Kaminski}, {Aceituno}, {Amado}, {B{\'e}jar}, {Hagen}, {Jeffers},
  {Quirrenbach}, {Launhardt}, {Marvin}, \& {Montes}}]{Sarkis18}
{Sarkis}, P., {Henning}, T., {K{\"u}rster}, M., {et~al.} 2018, \aj, 155, 257

\bibitem[{{Schweitzer} {et~al.}(2018){Schweitzer}, {Passegger}, \&
  {B{\'e}jar}}]{Schweitzer18}
{Schweitzer}, A., {Passegger}, V.~M., \& {B{\'e}jar}, V. J.~S. 2018, \aap, to
  be submitted

\bibitem[{{Skrutskie} {et~al.}(2006){Skrutskie}, {Cutri}, {Stiening},
  {Weinberg}, {Schneider}, {Carpenter}, {Beichman}, {Capps}, {Chester},
  {Elias}, {Huchra}, {Liebert}, {Lonsdale}, {Monet}, {Price}, {Seitzer},
  {Jarrett}, {Kirkpatrick}, {Gizis}, {Howard}, {Evans}, {Fowler}, {Fullmer},
  {Hurt}, {Light}, {Kopan}, {Marsh}, {McCallon}, {Tam}, {Van Dyk}, \&
  {Wheelock}}]{2MASS}
{Skrutskie}, M.~F., {Cutri}, R.~M., {Stiening}, R., {et~al.} 2006, \aj, 131,
  1163

\bibitem[{{Su{\'a}rez Mascare{\~n}o} {et~al.}(2017){Su{\'a}rez Mascare{\~n}o},
  {Gonz{\'a}lez Hern{\'a}ndez}, {Rebolo}, {Astudillo-Defru}, {Bonfils},
  {Bouchy}, {Delfosse}, {Forveille}, {Lovis}, {Mayor}, {Murgas}, {Pepe},
  {Santos}, {Udry}, {W{\"u}nsche}, \& {Velasco}}]{GJ536}
{Su{\'a}rez Mascare{\~n}o}, A., {Gonz{\'a}lez Hern{\'a}ndez}, J.~I., {Rebolo},
  R., {et~al.} 2017, \aap, 597, A108

\bibitem[{{Tal-Or} {et~al.}(2018){Tal-Or}, {Zechmeister}, {Reiners}, {Jeffers},
  {Sch{\"o}fer}, {Quirrenbach}, {Amado}, {Ribas}, {Caballero}, {Aceituno},
  {Bauer}, {B{\'e}jar}, {Czesla}, {Dreizler}, {Fuhrmeister}, {Hatzes},
  {Johnson}, {K{\"u}rster}, {Lafarga}, {Montes}, {Morales}, {Reffert},
  {Sadegi}, {Seifert}, \& {Shulyak}}]{Tal-Or18}
{Tal-Or}, L., {Zechmeister}, M., {Reiners}, A., {et~al.} 2018, \aap, 614, A122

\bibitem[{{Trifonov} {et~al.}(2018){Trifonov}, {K{\"u}rster}, {Zechmeister},
  {Tal-Or}, {Caballero}, {Quirrenbach}, {Amado}, {Ribas}, {Reiners}, {Reffert},
  {Dreizler}, {Hatzes}, {Kaminski}, {Launhardt}, {Henning}, {Montes},
  {B{\'e}jar}, {Mundt}, {Pavlov}, {Schmitt}, {Seifert}, {Morales}, {Nowak},
  {Jeffers}, {Rodr{\'{\i}}guez-L{\'o}pez}, {del Burgo}, {Anglada-Escud{\'e}},
  {L{\'o}pez-Santiago}, {Mathar}, {Ammler-von Eiff}, {Guenther}, {Barrado},
  {Gonz{\'a}lez Hern{\'a}ndez}, {Mancini}, {St{\"u}rmer}, {Abril}, {Aceituno},
  {Alonso-Floriano}, {Antona}, {Anwand-Heerwart}, {Arroyo-Torres}, {Azzaro},
  {Baroch}, {Bauer}, {Becerril}, {Ben{\'{\i}}tez}, {Berdi{\~n}as}, {Bergond},
  {Bl{\"u}mcke}, {Brinkm{\"o}ller}, {Cano}, {C{\'a}rdenas V{\'a}zquez},
  {Casal}, {Cifuentes}, {Claret}, {Colom{\'e}}, {Cort{\'e}s-Contreras},
  {Czesla}, {D{\'{\i}}ez-Alonso}, {Feiz}, {Fern{\'a}ndez}, {Ferro},
  {Fuhrmeister}, {Galad{\'{\i}}-Enr{\'{\i}}quez}, {Garcia-Piquer},
  {Garc{\'{\i}}a Vargas}, {Gesa}, {G{\'o}mez Galera}, {Gonz{\'a}lez-Peinado},
  {Gr{\"o}zinger}, {Grohnert}, {Gu{\`a}rdia}, {Guijarro}, {de Guindos},
  {Guti{\'e}rrez-Soto}, {Hagen}, {Hauschildt}, {Hedrosa}, {Helmling},
  {Hermelo}, {Hern{\'a}ndez Arab{\'{\i}}}, {Hern{\'a}ndez Casta{\~n}o},
  {Hern{\'a}ndez Hernando}, {Herrero}, {Huber}, {Huke}, {Johnson}, {de Juan},
  {Kim}, {Klein}, {Kl{\"u}ter}, {Klutsch}, {Lafarga}, {Lamp{\'o}n}, {Lara},
  {Laun}, {Lemke}, {Lenzen}, {L{\'o}pez del Fresno}, {L{\'o}pez-Gonz{\'a}lez},
  {L{\'o}pez-Puertas}, {L{\'o}pez Salas}, {Luque}, {Mag{\'a}n Madinabeitia},
  {Mall}, {Mandel}, {Marfil}, {Mar{\'{\i}}n Molina}, {Maroto Fern{\'a}ndez},
  {Mart{\'{\i}}n}, {Mart{\'{\i}}n-Ruiz}, {Marvin}, {Mirabet}, {Moya},
  {Moreno-Raya}, {Nagel}, {Naranjo}, {Nortmann}, {Ofir}, {Oreiro}, {Pall{\'e}},
  {Panduro}, {Pascual}, {Passegger}, {Pedraz}, {P{\'e}rez-Calpena}, {P{\'e}rez
  Medialdea}, {Perger}, {Perryman}, {Pluto}, {Rabaza}, {Ram{\'o}n}, {Rebolo},
  {Redondo}, {Reinhardt}, {Rhode}, {Rix}, {Rodler}, {Rodr{\'{\i}}guez},
  {Rodr{\'{\i}}guez Trinidad}, {Rohloff}, {Rosich}, {Sadegi},
  {S{\'a}nchez-Blanco}, {S{\'a}nchez Carrasco}, {S{\'a}nchez-L{\'o}pez},
  {Sanz-Forcada}, {Sarkis}, {Sarmiento}, {Sch{\"a}fer}, {Schiller},
  {Sch{\"o}fer}, {Schweitzer}, {Solano}, {Stahl}, {Strachan}, {Su{\'a}rez},
  {Tabernero}, {Tala}, {Tulloch}, {Veredas}, {Vico Linares}, {Vilardell},
  {Wagner}, {Winkler}, {Wolthoff}, {Xu}, {Yan}, \& {Zapatero
  Osorio}}]{Trifonov18}
{Trifonov}, T., {K{\"u}rster}, M., {Zechmeister}, M., {et~al.} 2018, \aap, 609,
  A117

\bibitem[{{Udry} {et~al.}(2007){Udry}, {Bonfils}, {Delfosse}, {Forveille},
  {Mayor}, {Perrier}, {Bouchy}, {Lovis}, {Pepe}, {Queloz}, \&
  {Bertaux}}]{Udry07}
{Udry}, S., {Bonfils}, X., {Delfosse}, X., {et~al.} 2007, \aap, 469, L43

\bibitem[{{Vanderburg} \& {Johnson}(2014)}]{K2}
{Vanderburg}, A. \& {Johnson}, J.~A. 2014, \pasp, 126, 948

\bibitem[{{West} {et~al.}(2015){West}, {Weisenburger}, {Irwin},
  {Berta-Thompson}, {Charbonneau}, {Dittmann}, \&
  {Pineda}}]{2015ApJ...812....3W}
{West}, A.~A., {Weisenburger}, K.~L., {Irwin}, J., {et~al.} 2015, \apj, 812, 3

\bibitem[{{Wright} {et~al.}(2016){Wright}, {Wittenmyer}, {Tinney}, {Bentley},
  \& {Zhao}}]{Wolf1061}
{Wright}, D.~J., {Wittenmyer}, R.~A., {Tinney}, C.~G., {Bentley}, J.~S., \&
  {Zhao}, J. 2016, \apjl, 817, L20

\bibitem[{{Wright} \& {Eastman}(2014)}]{WE14}
{Wright}, J.~T. \& {Eastman}, J.~D. 2014, \pasp, 126, 838

\bibitem[{{Zechmeister} \& {K{\"u}rster}(2009)}]{GLS}
{Zechmeister}, M. \& {K{\"u}rster}, M. 2009, \aap, 496, 577

\bibitem[{{Zechmeister} {et~al.}(2009){Zechmeister}, {K{\"u}rster}, \&
  {Endl}}]{Zechmeister09}
{Zechmeister}, M., {K{\"u}rster}, M., \& {Endl}, M. 2009, \aap, 505, 859

\bibitem[{{Zechmeister} {et~al.}(2018){Zechmeister}, {Reiners}, {Amado},
  {Azzaro}, {Bauer}, {B{\'e}jar}, {Caballero}, {Guenther}, {Hagen}, {Jeffers},
  {Kaminski}, {K{\"u}rster}, {Launhardt}, {Montes}, {Morales}, {Quirrenbach},
  {Reffert}, {Ribas}, {Seifert}, {Tal-Or}, \& {Wolthoff}}]{SERVAL}
{Zechmeister}, M., {Reiners}, A., {Amado}, P.~J., {et~al.} 2018, \aap, 609, A12

\end{thebibliography}


\begin{appendix} 

\section{Frequency-extended periodogram analysis} 

\begin{figure*}
\centering
\includegraphics[width=0.48\hsize]{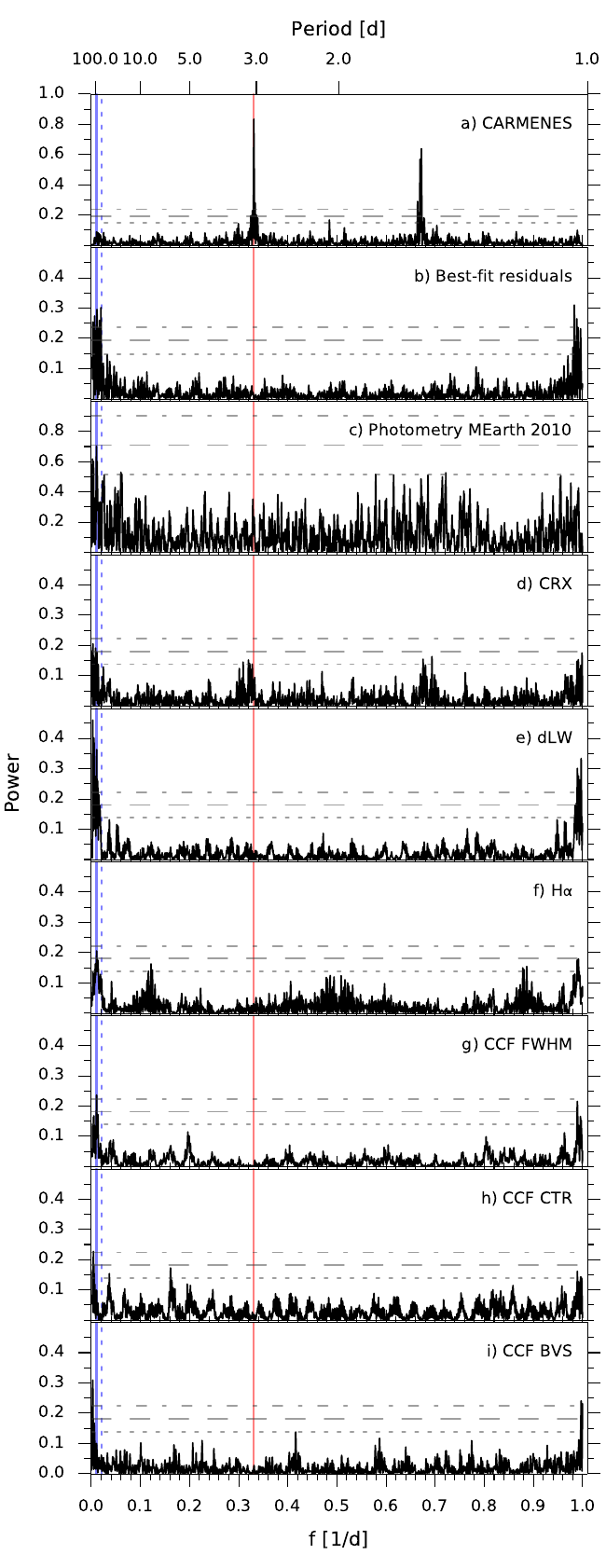}\ 
\includegraphics[width=0.48\hsize]{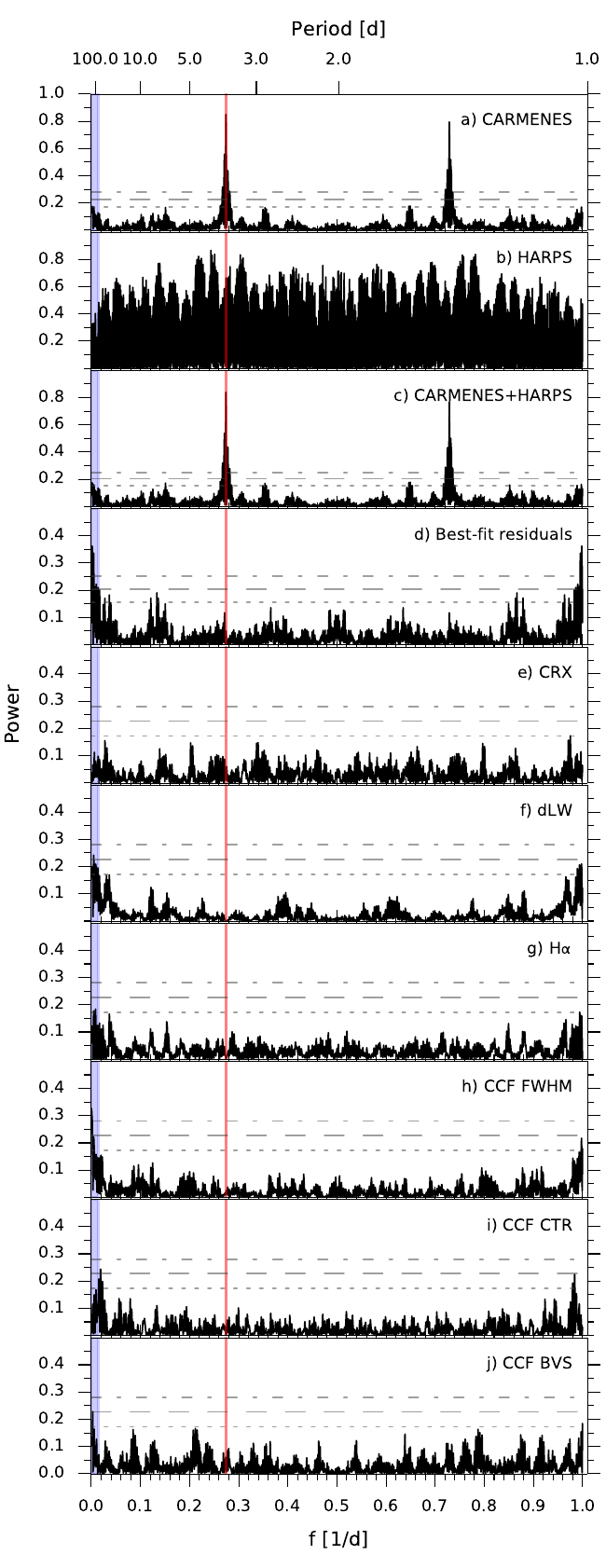}
\caption{Extended GLS periodograms for GJ~3779 ({\it left}) and GJ~1265 ({\it right}) covering the full frequency range. The magnitudes and the format represented are the same as in Figs.~\ref{fig:gj3779_gls} and \ref{fig:gj1265_gls}.}  \label{fig:gls_ext}
\end{figure*}

\section{Correlation between radial velocities and chromatic index} 

\begin{figure*}
\centering
\includegraphics[width=\hsize]{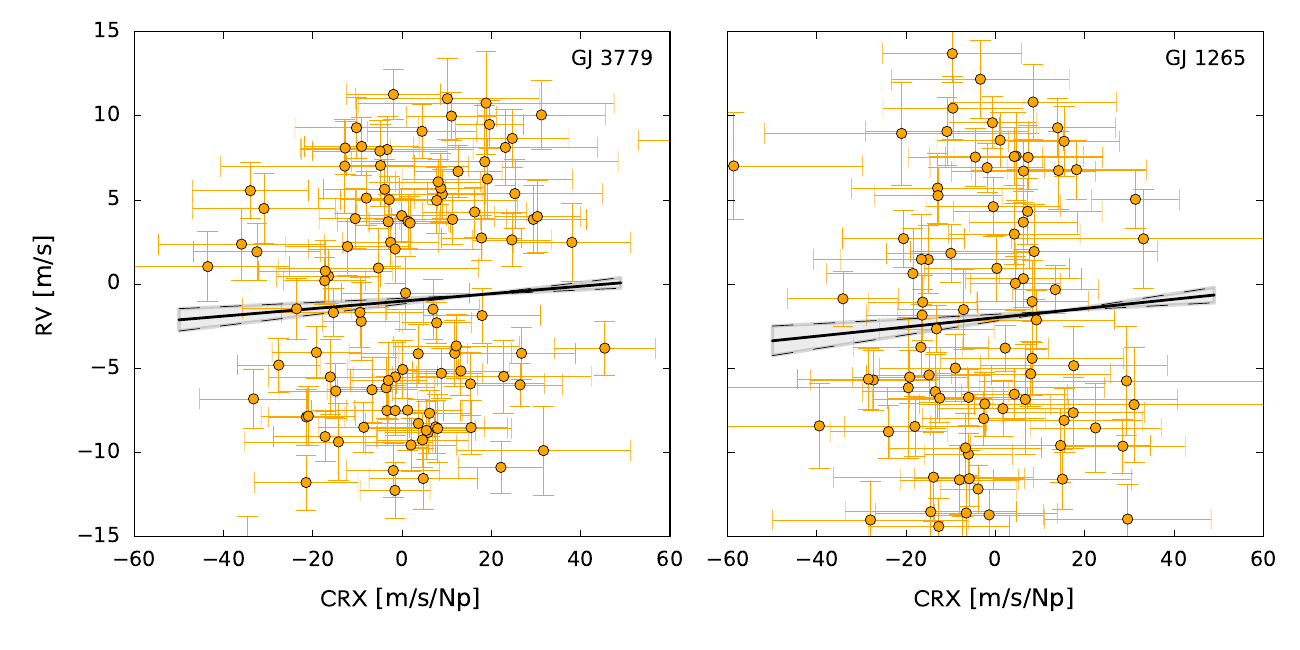}
\caption{CARMENES VIS radial velocities against the chromatic index (CRX) for GJ~3779 ({\it left}) and GJ~1265 ({\it right}). The slopes of each linear fit are consistent with zero within the errors. Besides, if there is activity-induced correlation, then the slope is usually negative \citep{Tal-Or18}.} \label{fig:rv_crx}
\end{figure*}

\section{MCMC analysis} 

\begin{figure*}
\centering
\includegraphics[width=\hsize]{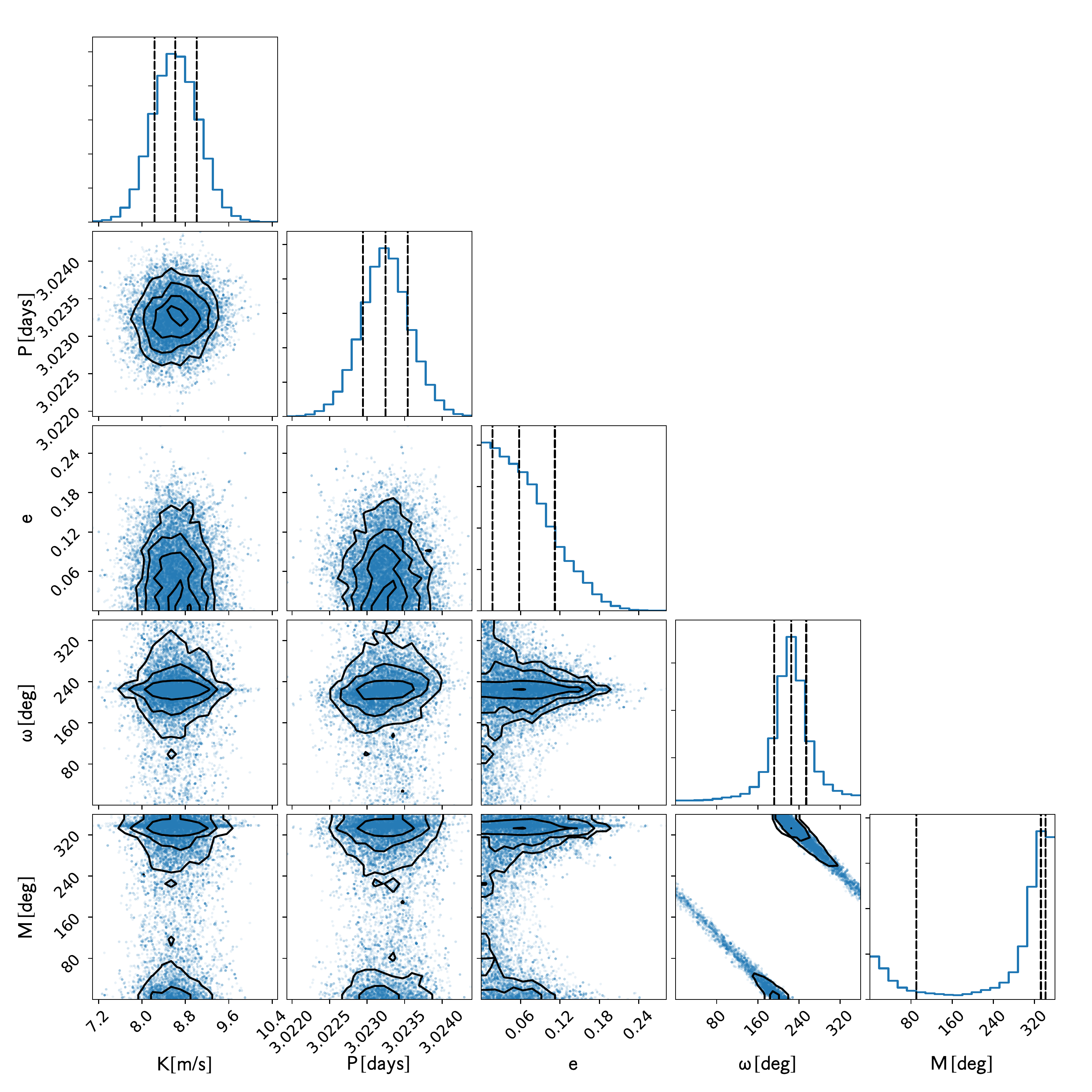}
\caption{MCMC results using the CARMENES RVs obtained for GJ~3779. Each panel contains approximately $40\,000$ Keplerian samples. The upper panels of the corner plot show the probability density distributions of each orbital parameter. The vertical dashed lines indicate the mean and 1$\sigma$ uncertainties of the fitted parameters. The rest of the panels show the dependencies between all the orbital elements in the parameter space. Black contours are drawn to improve the visualization of the two-dimensional histograms. The distributions agree within the errors with the best-fit values from Table~\ref{tab:fit_all}.
 } \label{fig:gj3779_mcmc}
\end{figure*}

\begin{figure*}
\centering
\includegraphics[width=\hsize]{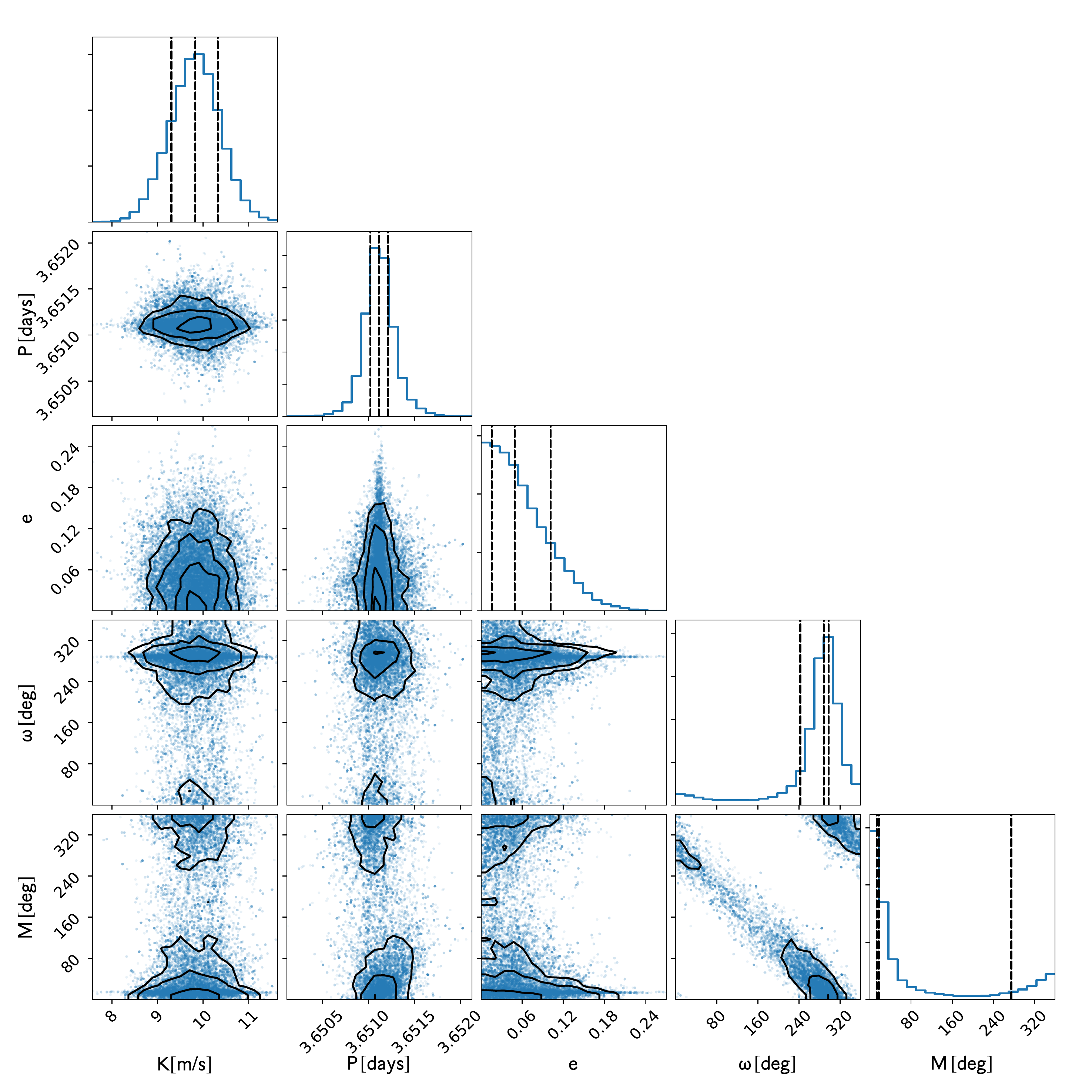}
\caption{MCMC results using CARMENES and HARPS datasets obtained for GJ~1265. Each panel contains approximately $40\,000$ Keplerian samples. The quantities represented and the format of the plot are the same as in Fig.~\ref{fig:gj3779_mcmc}. The distributions agree within the errors with the best-fit values from Table~\ref{tab:fit_all}.
 } \label{fig:gj1265_mcmc}
\end{figure*}

\newpage

\section{Radial velocity measurements}

\begin{table}
\caption{Radial velocities and formal uncertainties of GJ~3779.}\label{tab:gj3779_RVs}
{\renewcommand{\arraystretch}{1.0}
\begin{tabular}{lccr}
\hline\hline
\noalign{\smallskip}
$\mathrm{BJD}$  & RV                        & $\mathrm{\sigma_{RV}}$ & Instrument   \\
                & $[\mathrm{m\,s^{-1}}]$    & [$\mathrm{m\,s^{-1}}$] &              \\
\noalign{\smallskip}
\hline
\noalign{\smallskip}
2457489.5644 & -4.64 & 2.16 & CARMENES \\
2457499.5044 & 11.61 & 3.07 & CARMENES \\
2457509.5092 & 6.23 & 2.04 & CARMENES \\
2457509.5235 & 8.86 & 1.98 & CARMENES \\
2457531.4218 & -9.04 & 2.65 & CARMENES \\
2457564.4277 & -5.07 & 1.86 & CARMENES \\
2457802.6071 & 4.92 & 1.55 & CARMENES \\
2457823.5931 & 10.35 & 1.44 & CARMENES \\
2457852.5244 & -4.31 & 1.43 & CARMENES \\
2457855.5366 & -5.14 & 1.29 & CARMENES \\
2457861.5292 & -7.43 & 1.42 & CARMENES \\
2457876.4943 & -10.04 & 1.53 & CARMENES \\
2457880.4519 & 3.48 & 1.57 & CARMENES \\
2457893.5137 & 6.20 & 1.46 & CARMENES \\
2457897.3854 & -3.26 & 1.98 & CARMENES \\
2457901.3944 & 4.70 & 2.01 & CARMENES \\
2457904.4598 & 10.91 & 2.06 & CARMENES \\
2457907.3791 & 3.60 & 1.32 & CARMENES \\
2457907.4919 & 5.14 & 1.49 & CARMENES \\
2457909.4002 & -2.82 & 1.90 & CARMENES \\
2457909.5302 & -6.66 & 2.06 & CARMENES \\
2457910.3675 & 4.87 & 2.13 & CARMENES \\
2457910.4997 & 3.33 & 2.30 & CARMENES \\
2457911.4130 & 10.84 & 1.43 & CARMENES \\
2457912.3887 & -0.64 & 1.79 & CARMENES \\
2457912.5238 & -5.31 & 1.79 & CARMENES \\
2457914.4130 & 12.13 & 1.48 & CARMENES \\
2457915.3767 & 1.32 & 1.59 & CARMENES \\
2457915.5058 & -8.53 & 2.28 & CARMENES \\
2457916.4066 & 6.56 & 2.06 & CARMENES \\
2457918.4216 & -4.66 & 2.05 & CARMENES \\
2457919.4062 & -1.36 & 2.32 & CARMENES \\
2457920.5033 & 8.98 & 2.24 & CARMENES \\
2457921.4084 & -7.66 & 1.51 & CARMENES \\
2457922.4167 & 2.78 & 1.68 & CARMENES \\
2457924.4050 & -3.96 & 1.61 & CARMENES \\
2457924.4923 & -5.97 & 1.78 & CARMENES \\
2457964.3896 & -7.97 & 1.23 & CARMENES \\
2458095.6918 & 5.88 & 1.41 & CARMENES \\
2458110.7024 & 9.94 & 1.61 & CARMENES \\
2458112.7089 & -3.26 & 1.56 & CARMENES \\
2458117.7164 & -5.50 & 1.63 & CARMENES \\
2458121.7006 & -6.82 & 1.36 & CARMENES \\
2458122.6749 & 9.04 & 1.75 & CARMENES \\
2458123.7163 & -2.95 & 1.60 & CARMENES \\
2458134.6700 & 6.41 & 1.66 & CARMENES \\
2458135.6433 & 0.33 & 2.15 & CARMENES \\
2458136.6856 & -6.67 & 1.48 & CARMENES \\
2458139.7337 & -7.64 & 1.56 & CARMENES \\
2458140.5828 & 4.70 & 1.51 & CARMENES \\
2458140.7183 & 3.34 & 1.45 & CARMENES \\
2458141.6299 & -1.45 & 1.19 & CARMENES \\
2458142.6556 & -8.41 & 1.62 & CARMENES \\
2458149.5716 & 5.82 & 1.97 & CARMENES \\
2458161.6173 & 11.88 & 2.42 & CARMENES \\
2458166.7462 & -4.45 & 1.49 & CARMENES \\
2458172.6033 & -10.70 & 1.81 & CARMENES \\
2458172.6903 & -7.84 & 1.71 & CARMENES \\
\noalign{\smallskip}
\hline
\end{tabular}
}
\end{table}

\begin{table}
\caption{Table D.1 (cont.).}\label{tab:gj3779_RVs_cont}
{\renewcommand{\arraystretch}{1.0}
\begin{tabular}{lccr}
\hline\hline
\noalign{\smallskip}
$\mathrm{BJD}$  & RV                        & $\mathrm{\sigma_{RV}}$ & Instrument   \\
                & $[\mathrm{m\,s^{-1}}]$    & [$\mathrm{m\,s^{-1}}$] &              \\
\noalign{\smallskip}
\hline
\noalign{\smallskip}
2458200.6236 & 3.23 & 2.15 & CARMENES \\
2458205.5444 & -3.21 & 1.57 & CARMENES \\
2458206.6160 & 1.81 & 2.35 & CARMENES \\
2458209.5506 & -0.84 & 1.44 & CARMENES \\
2458212.5382 & 2.94 & 2.03 & CARMENES \\
2458213.5696 & 10.15 & 1.80 & CARMENES \\
2458215.4719 & -1.02 & 1.63 & CARMENES \\
2458236.4100 & -7.68 & 1.60 & CARMENES \\
2458237.4310 & 7.87 & 2.78 & CARMENES \\
2458238.5670 & -3.29 & 1.73 & CARMENES \\
2458244.5726 & -4.22 & 1.70 & CARMENES \\
2458245.5792 & -7.73 & 2.03 & CARMENES \\
2458247.6112 & -6.64 & 2.37 & CARMENES \\
2458260.4744 & -5.43 & 1.95 & CARMENES \\
2458261.3586 & 7.55 & 1.55 & CARMENES \\
2458261.5495 & 8.76 & 1.84 & CARMENES \\
2458262.5760 & 9.40 & 5.68 & CARMENES \\
2458263.3930 & -4.66 & 2.22 & CARMENES \\
2458264.3726 & 9.52 & 1.70 & CARMENES \\
2458264.4797 & 8.94 & 1.59 & CARMENES \\
2458265.4046 & 4.58 & 1.62 & CARMENES \\
2458269.5062 & -10.23 & 1.60 & CARMENES \\
2458270.4471 & 7.90 & 2.45 & CARMENES \\
2458271.4304 & 4.56 & 1.65 & CARMENES \\
2458273.4410 & 4.49 & 1.56 & CARMENES \\
2458277.4696 & 4.74 & 1.79 & CARMENES \\
2458280.4016 & 7.11 & 2.41 & CARMENES \\
2458284.3935 & -7.05 & 1.66 & CARMENES \\
2458284.4884 & -6.99 & 1.76 & CARMENES \\
2458289.3742 & 5.97 & 1.59 & CARMENES \\
2458290.4713 & -10.94 & 1.63 & CARMENES \\
2458291.3994 & 1.05 & 1.38 & CARMENES \\
2458292.3925 & 6.50 & 1.78 & CARMENES \\
2458293.3704 & -8.21 & 1.50 & CARMENES \\
2458294.4124 & 1.90 & 2.09 & CARMENES \\
2458295.3744 & 6.93 & 1.34 & CARMENES \\
2458296.4807 & -15.05 & 2.08 & CARMENES \\
2458297.4669 & 1.63 & 1.82 & CARMENES \\
2458299.3713 & -11.41 & 1.67 & CARMENES \\
2458300.4504 & -0.83 & 2.08 & CARMENES \\
2458301.4487 & 3.09 & 1.70 & CARMENES \\
2458304.4497 & 8.13 & 3.38 & CARMENES \\
2458305.3923 & -8.71 & 1.58 & CARMENES \\
2458306.4382 & -0.61 & 1.80 & CARMENES \\
2458307.4395 & 5.35 & 2.08 & CARMENES \\
2458309.3646 & -4.88 & 2.28 & CARMENES \\
\noalign{\smallskip}
\hline
\end{tabular}
}
\end{table}

\begin{table}
\caption{Radial velocities and formal uncertainties of GJ~1265.}\label{tab:gj1265_RVs}
{\renewcommand{\arraystretch}{1.0}
\begin{tabular}{lccr}
\hline\hline
\noalign{\smallskip}
$\mathrm{BJD}$  & RV                        & $\mathrm{\sigma_{RV}}$ & Instrument   \\
                & $[\mathrm{m\,s^{-1}}]$    & [$\mathrm{m\,s^{-1}}$] &              \\
\noalign{\smallskip}
\hline
\noalign{\smallskip}
2453203.8148 & 6.63 & 4.19 & HARPS \\
2453339.5473 & 13.82 & 3.10 & HARPS \\
2454052.5352 & 0.13 & 3.60 & HARPS \\
2454054.5746 & 11.73 & 3.14 & HARPS \\
2455517.5424 & 1.49 & 2.97 & HARPS \\
2456083.8528 & -14.76 & 6.33 & HARPS \\
2456084.9239 & 11.57 & 8.04 & HARPS \\
2456094.9279 & 22.31 & 10.07 & HARPS \\
2456096.9294 & 2.23 & 6.12 & HARPS \\
2456098.9380 & 9.55 & 5.83 & HARPS \\
2456119.8704 & -5.30 & 5.63 & HARPS \\
2457569.6589 & -7.17 & 2.62 & CARMENES \\
2457586.6559 & -0.79 & 2.20 & CARMENES \\
2457594.6092 & -11.46 & 2.03 & CARMENES \\
2457607.5732 & 13.19 & 1.73 & CARMENES \\
2457642.4868 & -6.64 & 1.73 & CARMENES \\
2457650.4998 & 2.69 & 1.80 & CARMENES \\
2457689.3382 & 2.41 & 1.35 & CARMENES \\
2457949.6208 & 0.14 & 2.06 & CARMENES \\
2457976.5635 & 11.32 & 2.23 & CARMENES \\
2457979.5600 & 9.09 & 1.76 & CARMENES \\
2457997.5121 & 3.24 & 1.56 & CARMENES \\
2457999.5468 & -6.12 & 1.58 & CARMENES \\
2458000.4715 & -8.30 & 2.15 & CARMENES \\
2458007.4525 & -9.53 & 1.60 & CARMENES \\
2458008.4627 & 4.27 & 1.45 & CARMENES \\
2458026.3722 & -0.87 & 1.66 & CARMENES \\
2458032.3912 & -2.98 & 2.06 & CARMENES \\
2458033.4096 & -6.56 & 2.06 & CARMENES \\
2458034.3806 & 13.74 & 2.16 & CARMENES \\
2458055.3079 & -3.49 & 1.59 & CARMENES \\
2458058.3441 & -7.09 & 1.45 & CARMENES \\
2458059.3263 & 1.91 & 1.30 & CARMENES \\
2458060.3258 & 6.08 & 2.03 & CARMENES \\
2458065.2734 & -2.05 & 1.63 & CARMENES \\
2458065.3012 & -0.56 & 1.65 & CARMENES \\
2458065.3241 & 1.06 & 1.81 & CARMENES \\
2458066.2825 & -8.91 & 2.06 & CARMENES \\
2458066.3077 & -9.04 & 2.02 & CARMENES \\
2458066.3313 & -9.89 & 2.40 & CARMENES \\
2458074.2868 & 5.03 & 1.74 & CARMENES \\
2458074.3421 & 3.92 & 1.65 & CARMENES \\
2458079.2639 & 4.27 & 1.62 & CARMENES \\
2458080.2675 & -8.84 & 1.46 & CARMENES \\
2458081.2525 & -1.88 & 1.63 & CARMENES \\
2458082.2639 & 9.78 & 1.74 & CARMENES \\
2458095.2594 & -5.00 & 1.25 & CARMENES \\
2458104.2516 & 12.47 & 3.15 & CARMENES \\
2458295.6553 & -2.78 & 1.97 & CARMENES \\
2458296.6583 & 5.03 & 2.83 & CARMENES \\
2458299.6339 & -4.34 & 2.56 & CARMENES \\
2458304.6624 & 11.82 & 2.58 & CARMENES \\
2458305.6540 & 7.90 & 1.82 & CARMENES \\
2458306.6497 & -2.50 & 3.34 & CARMENES \\
2458308.6492 & 13.91 & 1.56 & CARMENES \\
2458309.6633 & 1.47 & 2.41 & CARMENES \\
2458310.6382 & -7.39 & 2.21 & CARMENES \\
2458313.6678 & -4.34 & 1.52 & CARMENES \\
\noalign{\smallskip}
\hline
\end{tabular}
}
\end{table}

\begin{table}
\caption{Table D.3 (cont.).}\label{tab:gj1265_RVs_cont}
{\renewcommand{\arraystretch}{1.0}
\begin{tabular}{lccr}
\hline\hline
\noalign{\smallskip}
$\mathrm{BJD}$  & RV                        & $\mathrm{\sigma_{RV}}$ & Instrument   \\
                & $[\mathrm{m\,s^{-1}}]$    & [$\mathrm{m\,s^{-1}}$] &              \\
\noalign{\smallskip}
\hline
\noalign{\smallskip}
2458314.6416 & -3.96 & 4.18 & CARMENES \\
2458315.6388 & 11.16 & 3.06 & CARMENES \\
2458316.6249 & 7.91 & 2.09 & CARMENES \\
2458317.5689 & -3.55 & 3.25 & CARMENES \\
2458318.5972 & 4.14 & 2.83 & CARMENES \\
2458320.5882 & 4.90 & 2.96 & CARMENES \\
2458321.6134 & -4.65 & 3.26 & CARMENES \\
2458322.5516 & 4.91 & 1.70 & CARMENES \\
2458322.6343 & 7.46 & 1.70 & CARMENES \\
2458324.5823 & -4.92 & 1.90 & CARMENES \\
2458324.6571 & -7.53 & 2.08 & CARMENES \\
2458326.5556 & 9.01 & 1.77 & CARMENES \\
2458326.6511 & 8.92 & 1.87 & CARMENES \\
2458327.5609 & 4.04 & 1.71 & CARMENES \\
2458331.5362 & 3.68 & 2.66 & CARMENES \\
2458332.5745 & -7.43 & 1.62 & CARMENES \\
2458333.5527 & 9.79 & 1.72 & CARMENES \\
2458334.5385 & 11.28 & 2.03 & CARMENES \\
2458334.6273 & 9.74 & 1.92 & CARMENES \\
2458335.5159 & -4.20 & 2.20 & CARMENES \\
2458337.5800 & 12.67 & 1.88 & CARMENES \\
2458338.5832 & 6.53 & 1.97 & CARMENES \\
2458339.6139 & -2.79 & 1.99 & CARMENES \\
2458340.5036 & 5.18 & 1.80 & CARMENES \\
2458340.5650 & 3.14 & 1.95 & CARMENES \\
2458341.5601 & 14.39 & 2.31 & CARMENES \\
2458341.6501 & 13.03 & 2.22 & CARMENES \\
2458342.5160 & 1.33 & 1.64 & CARMENES \\
2458342.6283 & -3.33 & 1.57 & CARMENES \\
2458343.5361 & -5.79 & 1.61 & CARMENES \\
2458343.6212 & -3.13 & 1.92 & CARMENES \\
2458345.5215 & 9.76 & 1.71 & CARMENES \\
2458345.6206 & 9.12 & 1.47 & CARMENES \\
2458346.5006 & -4.58 & 1.46 & CARMENES \\
2458347.6224 & 2.83 & 2.69 & CARMENES \\
2458348.4860 & 10.76 & 1.64 & CARMENES \\
2458348.6300 & 15.90 & 1.70 & CARMENES \\
2458350.4669 & -6.26 & 1.78 & CARMENES \\
2458350.5852 & -6.23 & 2.53 & CARMENES \\
2458351.4821 & -3.44 & 1.80 & CARMENES \\
\noalign{\smallskip}
\hline
\end{tabular}
}
\end{table}

\end{appendix}

\end{document}